%% file: main.tex
\pdfoutput=1
\documentclass[12 pt]{elsarticle}
\usepackage{caption}
\usepackage{cancel}
\usepackage{amsmath}
\usepackage{amsfonts}
\usepackage{graphicx}
\usepackage[T1]{fontenc}
\usepackage[utf8]{inputenc}
\usepackage{rotating}
\usepackage{longtable}
\usepackage{booktabs}
\usepackage{dsfont}
\usepackage[margin=1.3in]{geometry}
\usepackage{subfig}
\usepackage{setspace}
\usepackage{appendix}
\usepackage{threeparttable}
\usepackage{hyperref}
\hypersetup{colorlinks=true,
            linkcolor=red,
            citecolor=black,
            filecolor=magenta,
            urlcolor=cyan,
            pdftitle={},
            pdfauthor={},
            pdfsubject={},
            pdfkeywords={},
            }
\usepackage{color}
\usepackage{lscape}
\usepackage{float}
\usepackage{umoline}
\usepackage{multirow}
\usepackage{fancyhdr}
\usepackage{chngcntr}
\usepackage{etoolbox}
\usepackage{lipsum}
\usepackage{caption}
\usepackage{pdfpages}
\usepackage{chronosys}
\usepackage{epigraph}

\usetikzlibrary{decorations.pathreplacing}
\makeatletter
\def\ps@pprintTitle{
  \let\@oddhead\@empty
  \let\@evenhead\@empty
  \def\@oddfoot{\reset@font\hfil\thepage\hfil}
  \let\@evenfoot\@oddfoot
}

\makeatother
\usepackage{titlesec}
\titleformat*{\section}{\Large\bfseries}
\titleformat*{\subsection}{\large\bfseries}
\titleformat*{\subsubsection}{\large\bfseries}
\titleformat*{\subparagraph}{\large\bfseries}
\titleformat{\paragraph}[runin]{\normalfont\normalsize\bfseries}{\theparagraph}{1em}{}

\biboptions{authoryear}
\begin{document}

\include{MS20230602-main-20251026}
\include{MS20230602-appendix-20251026}

\end{document}

%% file: MS20230602-main-20251026.tex
\begin{frontmatter}
\title{\Huge{\textbf{Media Slant is Contagious}}\tnoteref{tfoot}}\tnotetext[tfoot]{We thank for helpful comments and suggestions Julia Cagé, Mirko Draca, Ruben Durante, Claudio Ferraz, Roland Hodler, Eunji Kim, Michael Knaus, Ro'ee Levy, Josh McCrain, Giovanni Prarolo, Giacomo Ponzetto, Pia Raffler, Jim Snyder, Dmitriy Vorobyev, David Yang, Ekaterina Zhuravskaya, our anonymous referees, and seminar and conference participants at the Harvard Political Economy Seminar, Econometric Society World Congress, Bologna Workshop on Elections, Democracy, and Populism, Bolzano Workshop on Political Economy, Bergamo Advances in Economics Workshop, SIEP Conference at the University of L'Aquila, University of St.Gallen Economics Seminar, and Zurich Text as Data Workshop. Romina Jafarian, Wei Pin, and Matteo Pinna provided excellent research assistance. First version: December 2020. \protect\\ \indent \textit{Corresponding author:} Philine Widmer. \protect\\ \indent \textit{Email addresses:} \href{mailto:philine.widmer@unidistance.ch}{philine.widmer@unidistance.ch} (Philine Widmer), \href{mailto:cabedmeraim@princeton.edu}{cabedmeraim@princeton.edu} (Clémentine Abed Meraim), \href{mailto:sergio.galletta@uniroma1.it}{sergio.galletta@uniroma1.it} (Sergio Galletta), \href{mailto:ashe@ethz.ch}{ashe@ethz.ch} (Elliott Ash).
}
\author[UDS,PSE]{\large Philine Widmer}
\author[PRI]{\large{Clémentine Abed Meraim}}
\author[SAP]{\large{Sergio Galletta}}
\author[ETH]{\large Elliott Ash}
\address[UDS]{UniDistance Suisse}
\address[PSE]{Paris School of Economics}
\address[PRI]{Princeton University}
\address[SAP]{Sapienza University of Rome}
\address[ETH]{ETH Zürich}

\begin{abstract}
This paper examines the diffusion of media slant. We document the influence of Fox News Channel (FNC) on the partisan slant of local newspapers in the U.S. over the years 1995-2008. We measure the political slant of local newspapers by scaling the news article texts to Republicans' and Democrats' speeches in Congress. Using channel positioning as an instrument for viewership, we find that higher FNC viewership causes local newspapers to adopt more right-wing slant. The effect emerges gradually, only several years after FNC's introduction, mirroring the channel's growing influence on voting behaviour. A main driver of the shift in newspaper slant appears to be a change in local political preferences. 


\end{abstract}
\end{frontmatter}

\newpage

\section{Introduction}

It is well-documented that media outlets frame content in a way that is favourable to a particular political party or ideological perspective \citep[e.g.,][]{Groseclose_Milyo_2005,harmon2009generalsemantics,gentzkow2010mediaslant,puglisi2011newspaper}. These biases in reporting can impact public opinion \citep[e.g.,][]{Chiang_Knight_2011,djourelova2023persuasion}. In particular, there is extensive evidence that higher exposure to Fox News Channel boosts Republican vote shares and other conservative interests \citep{DellaVignaKaplan2007TQJoE,MartinYurukoglu2017AER,ash2023cable,ash2021effect}.

However, whether media slant is  contagious -- that is, whether biased news messaging spreads to other outlets -- remains largely unanswered. This issue is policy-relevant because media regulation in democracies typically aims at cultivating a diverse news market where consumers have an unrestricted choice among independent news sources. After all, diverse news sources only translate into diverse reporting if those sources are relatively independent in their newsmaking.

This paper studies such influence across media outlets in the context of U.S. cable TV news networks and local newspapers. Our focus is Fox News Channel (FNC), which has a well-documented conservative slant relative to other cable networks such as MSNBC and CNN. We ask whether higher FNC viewership among a local newspaper's readership leads to a change in the partisan slant of that newspaper's content. 

The first step in answering this question is measuring the slant of local newspaper content. The local news corpus comprises 15.1 million article snippets from 718 local newspapers in the United States covering the period from 2005 through 2008. %
Alongside the newspaper articles, we have close to 200,000 speeches from the U.S. Congressional Record over the same period.%
\footnote{To investigate the mechanisms, we also build an auxiliary corpus comprising newspaper articles from 1995 through 2004. There are an additional 31.4 million articles and almost 500,000 speeches from these earlier years. We consider this corpus as auxiliary because we only have data for fewer newspaper titles in those periods.}

We use our parallel corpora to construct a measure of local media slant, comparing the language in local newspaper articles to the language used in Congressional speeches. Methodologically, we follow the penalized multinomial model of speech proposed by \cite{gentzkow2019measuring}. We train a model based on the Congressional speeches to assess whether a given body of text more closely resembles the language used by Republicans rather than Democrats. We validate our model and find that Republican phrases emphasize right-wing priorities like tax reductions, fiscal conservatism, abortion issues, and immigration,\footnote{For example, the expression ``illegal alien'' emerges as predictive of Republican language \citep{djourelova2023persuasion}.} while Democratic phrases reflect left-wing priorities like social welfare, education, criticism of the Iraq War, and gun control. 
We then use the text model fitted in the Congressional Record to position each newspaper article on a left-right spectrum based on its text content. Higher (positive) values of that measure imply a Republican-leaning article, while lower (negative) values imply a Democratic-leaning article. We examine the most predictive phrases in newspaper articles to verify that our model, trained on Congressional speech patterns, effectively identifies political language in news coverage. The article-level measure is then aggregated to obtain a left-right slant measure at the newspaper level. For our main results, we calculate the share of Republican-leaning articles for each newspaper, and we show robustness to other aggregation rules.

Our main research question is whether newspaper content becomes more right-wing in response to higher FNC viewership in a newspaper's market (i.e., county). For causal estimates, we exploit exogenous variation in cable news exposure across counties coming from variation in the channel numbering of FNC. As first shown in \cite{MartinYurukoglu2017AER}, FNC's channel position is exogenous conditional on a set of geographical, demographic, and channel system-related controls \citep[see also][]{ash2023cable}. We document the instrument's first-stage relevance and provide evidence supporting the exogeneity assumption in our context. The FNC channel position is uncorrelated with historical (pre-FNC) newspaper slant and other local characteristics that predict viewership or slant.

We find that media slant is contagious: Higher local FNC viewership makes the slant of local newspapers more right-wing. According to our estimates, a one-standard-deviation increase in FNC viewership due to channel positioning increases right-wing newspaper slant by one-half a standard deviation. That effect magnitude corresponds to roughly one-tenth the Republican-Democrat difference in Congressional speeches and one-quarter the FNC-MSNBC difference in their news shows. Our results are robust to various checks, for example, regarding the slant-measure construction, instrument construction, and circulation weighting.

We then explore potential mechanisms for the Fox News effect on local newspapers. First, we explore a persuasion effect, i.e., FNC changing local policy preferences. We show that rightward shifts of newspaper slant coincide -- temporally and geographically -- with FNC's persuasion of voters \citep{DellaVignaKaplan2007TQJoE,MartinYurukoglu2017AER,ash2021effect}, who are presumably the readers of local newspapers. Regarding the timing of effects, newspaper slant changes several years after the introduction of FNC, mirroring the channel's growing influence on Republican vote shares over the same time. In terms of geography, we estimate heterogeneous treatment effects of FNC following the method of \cite{athey2019generalized} and show that counties with larger voting responses to FNC also have larger newspaper-slant responses. Persuasion could occur through FNC influencing the supply side (journalists, editors, and owners adopting more right-wing positions), the demand side (readers developing preferences for Republican-slanted content), or both. While we cannot cleanly disentangle supply from demand factors, we interpret the evidence as being particularly consistent with demand-side persuasion. FNC pushes the readership to the right, and local newspapers respond with more right-wing slant.\footnote{While journalists could be persuaded, too, it is unlikely that FNC's persuasion of voters would run through local newspapers only. \citet{ash2019conservative} make a similar point for judges: cable news influences criminal sentencing through voters and judicial elections rather than persuading judges directly.}

As a second possible mechanism, we examine cost-saving -- that is, local newspapers borrowing content from FNC rather than producing it themselves. The nationally oriented cable channels primarily cover news at the national or international level. Hence, this mechanism involves a potential trade-off if readers prefer local content (which is expensive to produce) over non-local content. Then, the local newspapers would borrow content as long as the savings outweigh the effect of losing some readers \citep[see][]{martin_mccrain_2019}. To evaluate whether direct borrowing is a key mechanism, we look for an FNC effect on the slant of local news topics -- e.g., events at local schools, hospitals, and government offices. Such local reporting constitutes the majority of the news articles in our sample. Since FNC does not feature any content on these topics, a rightward shift in local reporting would preclude cost-savings being a key driver of the main results. Indeed, we find a similar rightward shift for local-news topics. Thus, while newspapers likely directly borrow FNC content at times, our evidence suggests it is not a substantial contributor to the overall FNC effect on newspaper slant.

As a third and final mechanism, we discuss whether FNC's effect on local newspapers could work through competitive pressures in the local news market. Several suggestive analyses provide no evidence that competition drives the observed rightward shift in newspaper slant (e.g., we find no FNC effect on local newspaper circulation).

These findings add to the literature in economics and political science on biased media (e.g., \citeauthor{AshworthShotts2010JoPE}, \citeyear{AshworthShotts2010JoPE}; \citeauthor{Prat_2018}, \citeyear{Prat_2018}).\footnote{For surveys on the empirical and theoretical literature, see  \cite{Puglisi_Snyder_2015} and \cite{Gentzkow_et_al_2015}, respectively. See also \citet{Stroemberg_2015}.} This literature provides extensive evidence that mass media shift election outcomes and readers' political preferences. First, \cite{Gentzkow_Shapiro_2011} and \cite{drago2014meet} report that the opening of local newspapers boosts voter turnout. \cite{Chiang_Knight_2011} show that a newspaper endorsement for a presidential candidate shifts voting intentions in favour of this candidate. \cite{djourelova2023persuasion} shows, for the case of immigration and border security, that the language used in newspapers can causally shift readers' policy preferences. Beyond the United States, \cite{Enikolopov_et_al_2011} find that Russian voters with access to an independent television station are more supportive of anti-Putin parties.\footnote{Other prominent contributions on the mass media's persuasive effects around the world include \cite{adena2015radio}, \cite{dellavigna2014cross}, and \cite{yanagizawa2014propaganda}. \cite{prat_stroemberg_2013} and \cite{Stroemberg_2015} provide surveys on the mass media' political effects.}

Regarding FNC in particular, there is a body of research documenting its political and societal impact -- beyond shifting votes \citep{DellaVignaKaplan2007TQJoE,MartinYurukoglu2017AER,ash2021effect,li2022media}. It has also been shown that cable news can affect voter knowledge \citep{Hopkins_and_Ladd_2014,SchroederStone2015JoPE}, fiscal policy decisions \citep{ash2023cable}, as well as behaviours during the COVID-19 pandemic \citep{bursztyn2020misinformation,ash2020effect,Simonov2022}. %
We add to this work by looking at the influence of partisan narratives using text analysis, studying the spillover effects on other news outlets. 

Our main contribution to the debate on media bias lies in showing how news media outlets are interconnected. Recent contributions on cross-media influence document the influence of social media on traditional media \citep{cage2020social,hatte2021reading}, and that news outlets copy-paste from each other extensively \citep{cage2020production}. We document explicitly how \textit{media bias} from one media organization can causally spill over to other media organizations. Hence, our work identifies an additional potential channel through which partisan media affects political and social outcomes.

The diffusion of national partisan priorities into local news is important because local newspapers are pivotal for citizen engagement and political accountability \citep[e.g.,][]{Snyder_Stroemberg_2010}. \cite{GeorgeandWaldfogel2006} find that the market entry of a national media outlet (in their case, the New York Times) causes local outlets to focus more on local coverage. \cite{martin_mccrain_2019} show that the acquisition of U.S. local TV stations by the national conglomerate Sinclair leads to an increased share of national as opposed to local content. Further, \cite{Mastrorocco_Ornaghi_2020} document that these acquisitions by Sinclair reduce the local crime coverage and subsequently lower crime clearance rates. \cite{Angelucci2024} demonstrate that the introduction of television in the United States negatively impacted local newspapers' readership and advertising revenues, leading these newspapers to reduce their overall content, particularly local news, which in turn increased political alignment between congressional and presidential elections. We contribute to these debates by analysing how higher exposure to slanted national cable news changes local slant. Hence, our work adds to recent concerns about lower social cohesion when the role of local media as a moderating force is eroded \citep[e.g.,][]{djourelova2021impact,darr2018newspaper}.

Methodologically, our approach combines natural language processing (NLP), machine learning, and causal inference (see \citeauthor{gentzkow2019text} \citeyear{gentzkow2019text} and \citeauthor{ash2022text} \citeyear{ash2022text} for overviews of text-data-based work in economics). Regarding text-as-data approaches to measuring partisanship, the most related work is \cite{gentzkow2010mediaslant}, showing a correlation between local news slant and local partisan preferences \citep[see also][]{gentzkow2019measuring}.%

More broadly, our work contributes to the long-lasting debate on the importance of (un)biased media in democratic politics.%

\section{Data} \label{sec:data}

Our analyses combine text data with structured, county-level demographics and cable channel system information. Our text data comes from local newspaper articles and speeches in the Congressional record. The main observation period (2005-2008) is given by data availability, as we only have Nielsen viewership data for 2005-2008. In addition, we only have newspaper articles up until 2008. To run placebo checks and better understand the mechanisms behind our results, we also build slant measures for other periods between 1995 and 2008. %
In the following, we describe the sources of these different data types and their roles in the empirical analysis.

\paragraph{Local newspaper articles.}

The starting point is a corpus of local newspaper articles. Our source is the news aggregation site NewsLibrary, from which we obtain the headlines and article texts for a sample of local U.S. newspapers for 1995-2008.\footnote{Due to access limits from NewsLibrary, the available article text is capped at the first 80 words. This means that for longer articles, our texts only include the first 80 words. Full articles -- available on a pay-per-piece basis -- were prohibitively expensive given our coverage in time and space.} 
We programmatically read through the documents and extract the newspaper name, headline, article text, and publication date (for an example article, see Appendix Figure \ref{fig:newspap_example}). The corpus contains 46.6 million articles from 718 unique sources over the period 1995 through 2008. The main analysis for 2005-2008 uses 15.1 million articles. %
Appendix \ref{sec:app:data} provides more information on the sample.

For our analyses, we remove articles related to routine announcements, which include births, deaths, obituaries, and property notices. We remove those conservatively, using a dictionary with unambiguous terms like ``death notice''. Appendix \ref{sec:app:routine_announce} describes the filtering, and Appendix \ref{sec:app:slantmeasures} demonstrates robustness to not dropping those announcements.

\paragraph{Congressional speeches.}
To quantify the local newspaper articles' slant, we need a reference corpus on which we can build a slant model. For this, we use speeches from the digitized U.S. Congressional Record published by the Government Publishing Office \citep{gennaro2022emotion}. We collect about 670,000 speeches delivered by Republicans and Democrats (excluding Independents) for 1995-2008, with about 183,000 for the main-analysis period (2005-2008). See Appendix Table \ref{tab:congress_period} for additional information.

\paragraph{Newspaper-level circulation data.}
We match each local newspaper outlet with its primary county based on the newspaper name and the geographical information provided by NewsLibrary (e.g., \textit{The Call (Woonsocket, RI)} or the \textit{Albany Democrat-Herald (OR)}), the U.S. Newspaper Directory, or manual web searches. We then match each newspaper to the associated total circulation numbers (as of 2004) reported by \cite{Gentzkow_Shaprio_Sinkinson_2014}.

\paragraph{Channel positions and viewership.}
From Nielsen, we have yearly data on channel positions and ratings for Fox News Channel (FNC) and MSNBC for the years 2005-2008. These are the same data used, for example, by \cite{MartinYurukoglu2017AER} and \citet{ash2023cable}. First, we have the channel lineup for all U.S. broadcast operators and the respective areas served. Second, we have the viewership measure called ratings, which is proportional to the average number of minutes spent watching a channel per household. The channel lineup and viewership data come at the zipcode level, which we aggregate to the county level as the average across zipcodes weighted by the number of Nielsen survey respondents in the zipcode. After computing these values by county-year, we then collapse to the county level by averaging within county across 2005-2008.

\paragraph{Other demographic covariates.}

Finally, we have additional covariates, beginning with a rich set of demographics from the 2000 and 2010 censuses (see Appendix Table \ref{tab:sumstat}). For demographics, zipcode-level data are aggregated to the county level by averaging with weighting by zipcode population. We also use  Republican vote shares from 1996 and 2008, obtained from \citet{ash2021effect}.

\section{Measuring Media Slant} \label{sec:text_data_methods}

This section describes how we construct the main outcome variable in our regressions: a text-based measure of political slant in local newspapers. We aim to scale the texts of newspaper articles by their similarity to Republican (right-wing) and Democrat (left-wing) congressional speeches. To this end, we adapt the method of \cite{gentzkow2019measuring}.

Let $N$ be the set of congressional speeches indexed by $n$.  We build a model to predict the partisanship, $\text{Speech\_Partisanship}_n$,  for each congressional speech.  Positive values indicate a more Republican-leaning speech, while negative values indicate a more Democratic-leaning speech. Afterward, we will use the fitted model to label newspaper articles as resembling more closely Democrat or Republican speeches.  

\subsection{Text Pre-Processing and Featurization} \label{sec:text_data_methods_preprocessing}

First, we preprocess newspaper articles and congressional speeches. As noted by previous work on supervised learning with political speech \citep[e.g.,][]{denny2018text,PetersonSpirling2018PA}, the specific choices of pre-processing and featurization are not that critical and can be done for computational and interpretive convenience. Here, the preprocessing includes normalizing to lowercase, removing stopwords and non-informative elements (e.g., punctuation and formatting), tokenising the texts, and stemming the tokens. We then form bigrams (two-word phrases) from the stemmed tokens. Following previous work, we remove any bigrams that are mostly associated with congressional procedure, such as ``time question'' (\citeauthor{gentzkow2010mediaslant}, \citeyear{gentzkow2010mediaslant}; \citeauthor{gentzkow2019measuring}, \citeyear{gentzkow2019measuring}; \citeauthor{gennaro2022emotion}, \citeyear{gennaro2022emotion}). Appendix \ref{sec:app:preprocess} provides more details.

The starting point for the vocabulary is all bigrams that appear in both the speech corpus and the newspaper corpus. Further, we drop rare bigrams that would not be informative of partisanship.\footnote{We remove infrequent bigrams for efficient computation (in line with, e.g., \citeauthor{gentzkow2019measuring}, \citeyear{gentzkow2019measuring}, or \citeauthor{cagelepennecmougin2024}, \citeyear{cagelepennecmougin2024}).: To be included, a bigram must appear in at least 0.01\% of both Republican and Democrat speeches. Further, it must appear in at least 0.1\% of Republican speeches \emph{or} 0.1\% of Democrat speeches. We impose that bigrams must be in the newspaper corpus to ensure they capture broader political discourse in the media. See Appendix \ref{sec:app:freqfilter} for additional details on the vocabulary construction.} Let $V$ be the final vocabulary of 14,224 bigrams.

\subsection{Congressional Slant Model Estimation}\label{sec:text_data_methods:classify}

Our method for scaling slant with Congressional speeches is based on the penalized multinomial model of text by \cite{gentzkow2019measuring}. The multinomial inverse regression approach comes from \cite{taddy2013} and \cite{taddy2015}, with the penalized estimator approach coming from \cite{taddy2017} and \cite{gentzkow2019measuring}. A similar approach to ours is taken by  \cite{cagelepennecmougin2024}, who apply this estimator to scale left/right slant in campaign manifestos of French parliamentary candidates. We take a fitted model from one corpus (the speeches) to another corpus (the newspapers), similar to the cross-domain approach from \citet{osnabrugge2021cross}.

Formally, we assume that $c_{bn}$, the count of bigram $b$ in speech $n$, follows a multinomial distribution of the form
\[
c_{bn} \sim \text{MN}(q_{bn}, t_n)
\]
where $t_n$ is the total number of bigrams in speech $n$, and $q_{bn}$ is the probability of observing bigram $b$ in speech $n$. 
The probability of speech $n$ using bigram $b$ is given by
\[
q_{bn} = \frac{\exp(\alpha_b + \phi_b K_n)}{\sum_{b' \in V} \exp(\alpha_{b'} + \phi_{b'} K_n)}
\]
where $K_n$ is a binary indicator for party affiliation: $K_n = 1$ for Republican speeches and $K_n = 0$ for Democratic speeches. The parameter $\phi_b$ quantifies the strength of the association between usage of bigram $b$ and party affiliation, with more positive values reflecting a relative preference for $b$ among Republicans and more negative values a relative preference among Democrats. 

To estimate the parameters $\alpha_b$ and $\phi_b$, we deploy a distributed multinomial regression \citep{gentzkow2019measuring}. By approximating the multinomial likelihood with a Poisson distribution for $c_{bn}$, we derive the negative log-likelihood for bigram $b$ as:
\[
\ell(\alpha_b, \phi_b) = \sum_{n=1}^{N} \left[ t_n \exp(\alpha_b + \phi_b K_n) - c_{bn}(\alpha_b + \phi_b K_n) \right]
\]
$N$ is the total number of speeches in the dataset. To address overfitting, a lasso (L1) regularization term is added to the training objective. The penalized estimator is
\[
(\hat{\alpha}_b, \hat{\phi}_b) = \arg\min \left[ \ell(\alpha_b, \phi_b) + \lambda |\phi_b| \right]
\]
where $\lambda>0$ calibrates the strength of regularization. Following \cite{gentzkow2019measuring}, we choose the value of $\lambda$ that minimizes a Bayesian information criterion.\footnote{Given by $\text{BIC} = 2 \ell(\alpha_b, \phi_b) + \log(N)\, \text{df}$, where \(df\) is the degrees-of-freedom term, as described in \cite{Zou2007OnT}. This corresponds to the number of parameters with non-zero values after regularization, excluding \( \hat{\alpha}_b \), as outlined in \cite{taddy2015}.}

The slant estimator is fitted separately for the various time periods in the data that will play a role in our empirical analysis. The main model uses speeches from 2005-2008. The historical samples are 1995-1997, 1998-2000, and 2001-2004. These models are used, respectively, to score the slant of news articles for the same time period (see Appendix \ref{app:alt-model}).

\begin{figure}%
\centering
\caption{Democrat (Light Grey) and Republican (Dark Grey) Bigrams, 2005-2008 \label{fig:wordcloudcoef}}
\includegraphics[width=\textwidth]{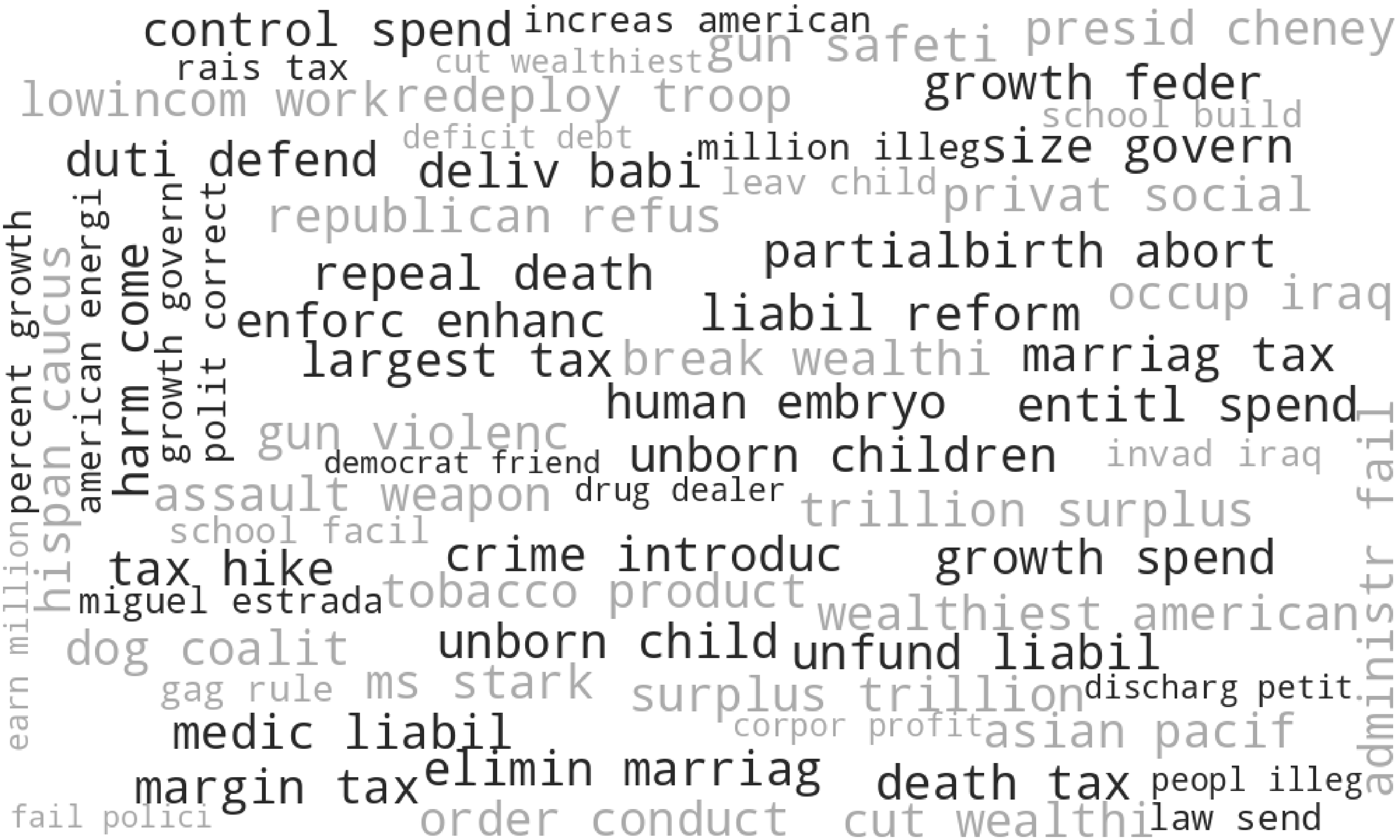}
\begin{minipage}{0.99\textwidth}
\vskip 0.5cm
{\footnotesize \emph{Notes:} The word cloud shows the most partisan bigrams for Democrats (light grey) and Republicans (dark grey) across newspapers during the period 2005--2008. We restrict the computation to the top 1,000 most partisan bigrams. Font size represents the relative partisanship of each bigram, with larger text indicating greater partisanship. Procedural bigrams have been filtered out. \par}
\end{minipage}

\end{figure}

The fitted model provides a set of partisan coefficients $\hat{\phi}_b$ for each bigram $b$. High values for $\hat{\phi}_b$ indicate right-wing bigrams and low values indicate left-wing. Figure \ref{fig:wordcloudcoef} presents a wordcloud of the most Republican (dark grey) and Democrat (light grey) bigrams for the main 2005-2008 sample, where the font size reflects the magnitude of the partisan coefficient $\hat{\phi}_b$. The results are intuitive. Republican discourse is focused on fiscal and social conservative priorities. Key bigrams like ``marriag tax,'' ``death tax,'' ``margin tax,'' and ``control spend'' reflect the Republican emphasis on tax reduction and fiscal restraint (recall that the bigrams were stemmed, hence this spelling). Social conservative priorities are evident in terms like ``partialbirth abort,'' ``unborn children,'' and ``human embryo,'' highlighting abortion-related issues. Meanwhile, Democratic discourse covers social welfare and international affairs. Bigrams like ``lowincom work,'' ``cut wealthi,'' and ``school facil'' suggest a focus on social programmes, economic equality, and education funding. Terms like ``redeploy troop,'' ``occup iraq,'' and ``invad iraq'' reflect Democratic criticism of Republican foreign policy, particularly regarding the Iraq War. The presence of ``gun safeti'' and ``gun violenc'' indicates Democratic emphasis on firearm regulation.

Appendix Figure \ref{fig:wordcloudfreq} plots the most partisan bigrams but with the font size proportional to the bigram frequency in newspapers. This figure illustrates that in the context of the newspapers, Republican slant is mostly around conservative fiscal policy, immigration (e.g., ``illeg alien''), and abortion. Democrat slant centres around social and health policy issues. 

\subsection{Measuring Slant in Newspapers} \label{subsec:text-outcome}

The next step is to use the model estimates $\hat{\phi}_b$ for each bigram to compute the newspaper articles' partisanship score. The partisanship of article $m$ in newspaper $i$ is the weighted sum of the coefficients across the bigrams appearing in the article:
\begin{equation} \label{eq:articleslant}
\text{Article\_Partisanship}_{{mi}}
 = \sum_{b \in V} f_{bmi} \cdot \hat{\phi}_b
\end{equation}
where $f_{bmi}$ is the relative frequency of bigram $b$ in article $m$.\footnote{Specifically, the relative frequency is given by the count of bigram $b$ in article $m$, divided by the total number of bigrams in article $m$ from the vocabulary $V$.}

Positive values of $\text{Article\_Partisanship}_{mi}$ indicate that an article leans Republican. Negative values imply it is Democratic-leaning. Scores near zero reflect mixed or neutral language. To focus on the presence of right-wing slant, we convert the continuous score to a discrete classifier for ``is right wing''. Specifically, we define $\text{Rep}_{mi}$ = 1 for $\text{Article\_Partisanship}_{mi}>0$ and $\text{Rep}_{mi}$ = 0 otherwise.

Finally, for our regressions, we aggregate up to the newspaper level. The main outcome for newspaper $i$ is the share of articles classified as Republican:
\begin{equation} \label{eq:outletslant}
\text{Slant}_i =  \frac{1}{M_i} \sum_{m=1}^{M_i} \text{Rep}_{mi} 
\end{equation}
where $M_i$ is the number of articles produced by newspaper $i$. In the remainder of the paper, when referring to the newspapers' ``slant,'' we mean $\text{Slant}_i$ as defined in Equation \ref{eq:outletslant} unless otherwise specified. For robustness, we compute alternative slant measures: (i) the share of Republican articles among all \textit{partisan} articles (i.e., excluding articles with $\text{Article\_Partisanship}_{mi}=0$), (ii) the newspaper-level average of the partisanship scores described by $\text{Article\_Partisanship}_{mi}$ in equation \ref{eq:articleslant}, and (iii) the share of Democrat articles (i.e., the same as in Equation \ref{eq:outletslant} but for Democrat articles). We discuss the results for these additional measures in the robustness checks (Section \ref{sec:robust} and Appendix \ref{sec:app:slantmeasures}). Appendices \ref{app:bigrams} to \ref{sec:app:examplearticles} further describe and validate the slant measure.

\subsection{Additional Text Variables}

As part of the analysis below, we use several other variables constructed from the text. First, we have some news article controls to capture the non-ideological style features. These include the unique-to-total word count ratio, average word length, and average article length.

Next, when studying the mechanisms in Section \ref{sec:mech}, we need to classify news articles as either referring to local news topics or higher-level news (i.e., national or international news). To do this, we first label 100,000 random newspaper article snippets as either ``local'' or ``non-local'' using GPT-3.5-Turbo. Then, we train a supervised binary classifier on these labelled data points. The classifier is then used to label all newspaper articles as local vs. non-local. Appendix \ref{sec:app:local_newspaper} explains this procedure.

\section{Econometric Framework} \label{sec:econometrics}

\subsection{Instrumental Variables Specification}

The main outcome variable is $\text{Slant}_{i}$, the slant of newspaper $i$ circulating in county $j$ (see Section \ref{subsec:text-outcome} above). Higher values of $\text{Slant}_{i}$ indicate that a larger share of articles from newspaper $i$ resemble Republican speeches. We are interested in the causal effect of local FNC viewership on slant. Hence, our main treatment variable, $\text{Viewership}_{j}$, is the county-level FNC viewership.\footnote{In the robustness checks, we also look at FNC relative to MSNBC channel viewership, see Appendix \ref{sec:app:fncmsnbc}.}  We specify the relationship between slant and viewership linearly:
\begin{equation}\label{eq:ols}
\text{Slant}_{i} =\alpha_s +\theta \text{Viewership}_{j} + \mathbf{X1_{j}}  \beta_1 + \mathbf{X2_{i}} \beta_2 + \epsilon_{i}
\end{equation}
where $\theta$ is the causal parameter of interest. The regression includes state fixed effects ($\alpha_s$), a vector of county-level controls, $\mathbf{X1_{j}}$, and a vector of newspaper-level controls, $\mathbf{X2_{i}}$. The error term is $\epsilon_{i}$. The exact covariate set $\mathbf{X1_{j}}$ varies across specifications. It always includes demographic controls and can additionally contain channel controls: the share of households with potential access to FNC, CNN, or MSNBC and the position of MSNBC and CNN.\footnote{These controls follow, as closely as possible, the channel controls used by \cite{MartinYurukoglu2017AER}. Some of Martin and Yurukoglu's controls are rendered moot with our county-level observations since they do not vary much across counties in 2005-2008. This applies to the total number of channels and the number of broadcast channels on the system.} Appendix Table \ref{tab:sumstat} provides details. Newspaper-level controls $\mathbf{X2_{i}}$ capture generic language controls (unique-to-total word count ratio, average word length, average article length).

Estimating Equation (\ref{eq:ols}) using OLS produces biased $\theta$ estimates. For a discussion of the OLS estimates, see Appendix \ref{sec:app:ols}. We use an instrumental variable to produce causal estimates. Following \cite{MartinYurukoglu2017AER}, we use cable network channel positioning to construct an instrument $\text{Position}_{j}$ that affects $\text{Viewership}_{j}$ but is otherwise unrelated to any factors affecting $\text{Slant}_{i}$. As first shown by \cite{MartinYurukoglu2017AER} and since used in various papers \citep[e.g.,][]{ash2023cable}, there is arbitrary variation in cable channel positioning across U.S. localities. This channel positioning leads to exogenous shifts in viewership because television watchers spend more time on networks with lower channel positions. Thus, we instrument viewership using channel position.\footnote{We also report results for the MSNBC channel position, which comes with similar quasi-experimental variation, see Appendix \ref{sec:app:msnbc}.} The first stage is:

\begin{equation}\label{eq:fs}
\text{Viewership}_{j} =\alpha_s + \delta \text{Position}_{j} + \mathbf{X1_{j}}  \beta_1 + \mathbf{X2_{i}} \beta_2 + \eta_{i}
\end{equation}

Combining the first stage (\ref{eq:fs}) and Equation (\ref{eq:ols}), we can obtain causal estimates for the local average treatment effect $\theta$ using two-stage least squares (2SLS). To facilitate the coefficient interpretation, we standardise the instrument, endogenous regressor, and outcome by dividing the original values by the standard deviations. Standard errors are clustered by state or, for robustness checks, by census division or without clustering. Our baseline regression weights newspapers by circulation, with robustness checks on weighting reported in Appendix \ref{sec:app:weights}.%

\subsection{Instrument First Stage and Validity}

\begin{figure}%
\centering
\caption{First Stage: Fox News Viewership and Channel Position} \label{fig:first-stage}
{\includegraphics[width=0.99\textwidth, keepaspectratio]{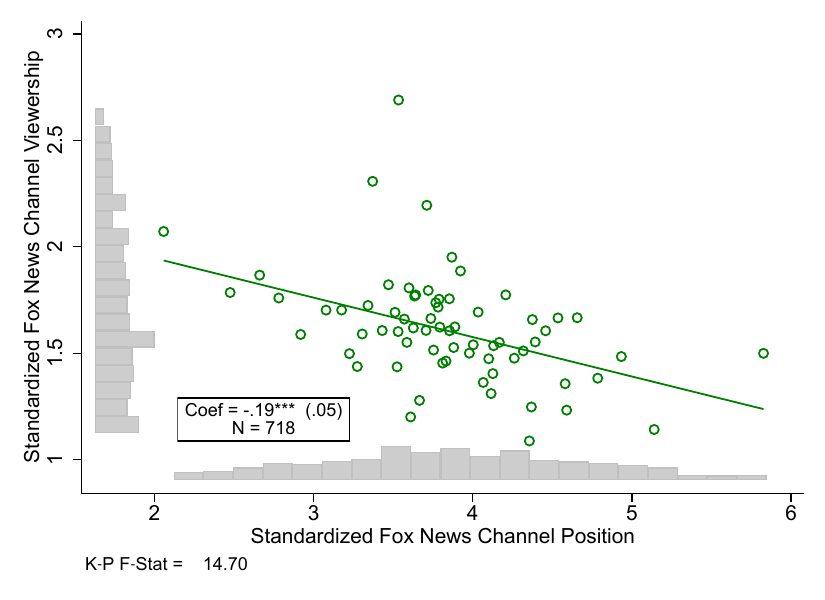}}
\begin{minipage}{0.99\textwidth}
\vskip 0.2cm
{\footnotesize \emph{Notes:} Binned scatterplots (equally sized bins) of the standardised FNC viewership (vertical axis) against the standardised FNC channel position (horizontal axis). Cross-section with newspaper-level observations weighted by newspaper circulation. We include state fixed effects, demographic controls, and channel controls  (see Table \ref{tab:sumstat} for the specific controls). Next to the axes, we show the distributions of the underlying variables in light grey. The reported coefficients and standard errors are from OLS estimates of the first stage with clustering by state, along with the observation count and the Kleibergen-Paap F-Statistic.\par}
\end{minipage}
\end{figure}

Figure \ref{fig:first-stage} visualizes the first-stage relation between the FNC channel position (horizontal axis) and FNC viewership (vertical axis) in each newspaper's county. The figure includes state fixed effects, channel controls, and demographic controls. Consistent with the logic of the instrument and the previous literature, the relationship is significantly negative. A one-standard-deviation decrease in FNC's channel position (11 positions in the lineup) increases the channel's viewership by about 19\% of a standard deviation (roughly 0.05 rating points). Our first-stage coefficient means that decreasing the FNC channel position by one standard deviation or 11 positions increases the channel's viewership by roughly 23 minutes per month.\footnote{One-tenth of a rating point equals 45 minutes per month of (additional) viewership per household.}

Appendix Table \ref{tab:first-stage} shows coefficients and standard errors in tabular format. The table's upper panel shows different specifications for the association of the FNC position and FNC viewership (adding different fixed effects and controls). Census division fixed effects (which group several states) in columns (1) to (3) and state fixed effects in columns (4) to (6) lead to similar estimates. We thus concentrate on the discussion of columns with state fixed effects. In line with \cite{MartinYurukoglu2017AER}, we find a strong first stage once we include demographic controls (moving from column 4 to 5) and especially when additionally introducing channel controls (moving from column 5 to 6). Column (6) is our preferred specification (also shown in Figure \ref{fig:first-stage}). Our instrument is relevant, with a Kleibergen-Paap cluster-robust first-stage F-statistic of 15.
\footnote{For completeness, Appendix \ref{sec:app:fs} includes complementary analyses of the first stage with MSNBC viewership and positioning.} %

Beyond relevance, we assume monotonicity, that the exclusion restriction holds, and exogeneity. Monotonicity appears plausible in our context (the channel position influences TV viewership in the same direction for all counties, and thus, higher positions do not systematically increase viewership). For the exclusion restriction, we assume that the channel position affects local newspapers only through its effect on cable news viewership. Exogeneity requires that $\text{Position}_{j}$ is uncorrelated with $\epsilon_{i}$ (conditional on controls, see \citeauthor{MartinYurukoglu2017AER}, \citeyear{MartinYurukoglu2017AER}. That is, the channel position is not endogenously selected with county-specific preferences for conservative or liberal news reporting. The main identification problem is that channel positions could be allocated endogenously in response to local factors correlated with conservative news messaging. 

\cite{MartinYurukoglu2017AER} provide a detailed discussion and several checks supporting the exogeneity assumption. Their qualitative research highlights that channel positions have an important arbitrary, historical component with significant inertia and path dependence. Quantitatively, they document that the instrument is uncorrelated with Republican vote shares before the introduction of FNC. They also show that the instrument is unrelated to demographic characteristics that predict policy preferences or news channel viewership. \cite{ash2023cable} report similar reassuring checks in a county-level regression specification. We apply the same checks to our newspaper-level data, finding no association between the FNC channel position and county characteristics otherwise important for our endogenous regressor or outcome (Appendix \ref{sec:app:exogeneity}). Furthermore, in Figure \ref{fig:timedyn}, we will look at the timing of the effect (also including a placebo outcome of local news slant from 1995 and 1996, before the introduction of FNC). We find no placebo relationship and show that the effect emerges over time. Accordingly, pre-existing newspaper characteristics do not drive our results. %

\section{Results} 
\label{sec:results:2sls}

\subsection{Main Results} \label{sec:main_results}

\begin{figure}%
\centering
\caption{Reduced Form: Republican Newspaper Slant and Fox News Position \label{fig:reduced-form}} 
{\includegraphics[width=0.99\textwidth, keepaspectratio]{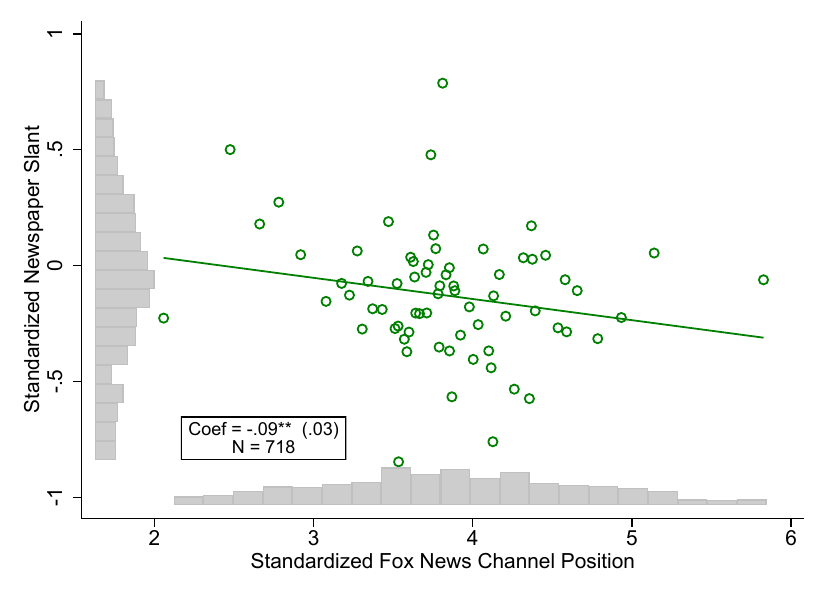}}
\begin{minipage}{0.99\textwidth}
\vskip 0.2cm
{\footnotesize \emph{Notes:} Binned scatterplots (equally sized bins) of the standardised newspaper slant (vertical axis) against the standardised FNC channel position (horizontal axis). Cross-section with newspaper-level observations weighted by newspaper circulation. We include state fixed effects, demographic controls, and channel controls (see Table \ref{tab:sumstat} for the specific controls). Next to the axes, we show the distributions of the underlying variables in light grey. The reported coefficients and standard errors are from OLS estimates of the reduced form with clustering by state, along with the observation count. \par}
\end{minipage}
\end{figure}

Figure \ref{fig:reduced-form} shows the reduced form results. The specification is the same as Figure \ref{fig:first-stage}, with standardised FNC channel position on the horizontal axis, but now with the main outcome variable, standardised newspaper slant, on the vertical axis. There is a clear negative relationship that is statistically significant in the associated regression. This is our first causal evidence that media slant is contagious. If the channel position of FNC in a newspaper's county is one standard deviation lower, slant is roughly one-tenth of a standard deviation more Republican.

Table \ref{tab:main_instr} shows two-stage-least-squares estimates for the effect of higher FNC viewership on the slant of local newspapers. The right-hand side variable of interest is instrumented FNC viewership. All columns include state fixed effects and demographic controls. Column (2) also includes channel controls. Column (3) additionally controls for generic newspaper language features. Both the explanatory and the outcome variables are standardised.

\input{tables/main_instr}

In all three columns, the estimated treatment effects are positive and statistically significant. The magnitudes are 0.72 in column (1), 0.49 in column (2), and 0.52 in column (3). Thus, while the channel controls reduce the coefficient size, the newspaper language controls leave the estimate largely unchanged. The differences across the three columns are safely within the confidence intervals. The interpretation is as follows: If FNC viewership increases by one standard deviation in the county $j$ where newspaper $i$ circulates, $i$'s content becomes more slanted towards Republican language by about half of a standard deviation.\footnote{OLS estimates are reported in Appendix \ref{sec:app:ols}, and Appendix \ref{sec:app:rf} shows all reduced form results in tabular form. We typically do not find significant effects of instrumented MSNBC viewership on slant (see Appendix \ref{sec:app:msnbc}).}

Since our 2SLS estimation gives the local average treatment effect (LATE), it captures the effect for newspapers responsive to the channel-position instrument. Our 2SLS estimates are larger than the OLS estimates (in line with \citeauthor{MartinYurukoglu2017AER}, \citeyear{MartinYurukoglu2017AER}). Measurement error is one likely explanation for this, as discussed in Appendix \ref{sec:app:ols}. Furthermore, the compliers may experience larger effects than the average newspaper, analogously to the argument in \citeauthor{MartinYurukoglu2017AER} (\citeyear{MartinYurukoglu2017AER}, p. 2579) that less partisan voters may be more persuadable by an exogenous increase in FNC exposure. In our case, this argument is plausible both from a demand and supply viewpoint. Newspapers serving counties with persuadable readers plausibly shift their slant more than newspapers serving those with stronger partisan preconceptions. On the supply side, it is equally plausible that complier editors, journalists, or owners are those with weaker pre-existing partisan agendas. %

Our effects are larger than what back-of-the-envelope calculations based on previous work would suggest: \cite{MartinYurukoglu2017AER} estimate that Fox News increased the Republican vote share by 6.34 percentage points in 2008, while \cite{gentzkow2010mediaslant} find that a 10 percentage point increase in Republican vote share correlates with a 0.375 standard deviation increase in newspaper slant. Chaining these estimates together predicts a Fox News effect of roughly 0.24 standard deviations (6.34 $\times$ 0.375/10 $=$ 0.24). Several possible explanations may account for this difference. Our instrumental variable approach captures the effects of complier newspapers (see above), contrary to the correlation from \cite{gentzkow2010mediaslant}. For a given shift in reader preferences, these newspapers may respond particularly strongly (e.g., contrary to newspapers whose owners are more strongly ideological). Further, their correlation reflects a long-run equilibrium in which reader preferences and slant co-evolve over long time periods. We, meanwhile, describe the effects of a shock (i.e., the rapid expansion of FNC). %
Of course, slight differences in slant measures, time periods, and newspaper samples may also idiosyncratically explain the differences in magnitudes.

Next, to help interpret our magnitudes, let us compare our effect with slant differences in Congressional speeches. For this, we apply our model to Congressional speeches in 2005-2008 (the corpus on which it was trained) and predict the average slant of Republican and Democrat speeches. The respective values are 0.71 and 0.42 -- i.e., 71\% of Republican speeches are classified as right-wing by the model; 42\% of Democrat speeches are classified as right-wing. The difference between the two parties is 0.29 (see details in Appendix \ref{sec:app:validation}). The standard deviation of slant in our newspaper sample is 0.07. Thus, if FNC viewership increases by one standard deviation, newspapers move to the right by slightly more than one-tenth of the difference between the average Republican and Democrat partisanship in 2005-2008. For a comparison from the media sector, let us also scale the effect to the reporting of FNC vs. MSNBC. The slant values are 0.48 and 0.34, respectively, giving a difference of 0.14. Our effect size is thus about one-quarter of the difference between the two channels at that time.\footnote{In the Congressional Record, less than 1\% of speeches are predicted as neutral (with a score of exactly zero). This may in part be due to the model being trained on that corpus; fewer speeches will not contain any predictive bigram and hence obtain a zero score by construction. Meanwhile, 17\% of the cable transcripts are coded as neutral. If we focus only on \textit{partisan} TV transcripts, the difference between the two channels is more similar to the Republican-Democrat difference in Congress (now 0.58 vs. 0.40). Our results for newspapers are robust to only focusing on partisan articles (see the robustness checks right below).}

\subsection{Robustness Checks} \label{sec:robust}

First, we present robustness checks for the slant measure. In Table \ref{tab:raw}, we average all articles' raw slant scores (i.e., without first dichotomizing them into Republican and Democrat articles) to obtain the newspaper-level slant. Qualitatively similar results emerge. Appendix \ref{sec:app:slantmeasures} shows additional slant measurement-related checks. For example, our results are very similar when measuring newspaper slant as the share of Republican articles in partisan articles (i.e., Democrat or Republican articles) instead of the share among all articles.  %

\input{tables/main_proba}

Second, we look at sampling and weighting. We confirm robustness to excluding outlets with sparse coverage ($<$1000 articles per year, which could generate noisy slant measures) and including routine announcements (e.g., obituaries). Next, we look at different weighting schemes in Appendix \ref{sec:app:weights}. %
Our results remain similar. Third, in Appendix \ref{sec:app:fncmsnbc}, we define the instrument differently. We find qualitatively equivalent effects when using FNC viewership relative to MSNBC viewership (instrumented by the difference between the position of the two channels). Fourth, Appendix \ref{sec:app:misc} shows miscellaneous checks, such as robustness to using census division (instead of state) fixed effects and dropping observations where viewership is likely measured imprecisely (low number of Nielsen respondents). We also run leave-one-out analyses at the state level. Our findings are not driven by any single state and, by extension, any single newspaper.
 
\section{Mechanisms} \label{sec:mech}

An important question regarding slant contagion is \textit{why} newspaper content becomes more conservative in response to higher FNC viewership. Among the explanations that could (jointly) play a role are (i) persuasion, (ii) cost-saving motives, and (iii) market competition.

\paragraph{Persuasion.}
Both supply-side and demand-side actors could be persuaded by FNC.
On the supply side, coverage by FNC could persuade journalists and editors, leading them to make reporting and framing choices in line with that channel's content (or make previously conservative journalists and editors perceive such reporting as more professionally acceptable). Moreover, it could persuade owners to change their editorial stance. On the demand side, FNC could directly influence readers' preferences for more conservative content. In our setting, we cannot cleanly disentangle supply from demand factors.\footnote{The literature provides mixed evidence. Some work shows that slant across U.S. newspapers reflects the political preferences of readers rather than producers \citep{gentzkow2010mediaslant}, while other work documents that media ownership influences or is correlated with news content \citep{Gilens_2000,martin_mccrain_2019,Mastrorocco_Ornaghi_2020,Szeidl_Szucs_2021,matter2021owns}.} In what follows, we show evidence that slant changes seem to coincide with changes in voting, which could be more supportive of the demand side: While FNC potentially influences journalists, it is unlikely that FNC's impact on voters would occur solely through local newspaper coverage.

\begin{figure}%
\centering
\caption{\centering Reduced Form: (Historical) Republican Newspaper Slant and Fox News Position} \label{fig:timedyn}
{\includegraphics[width=0.99\textwidth, keepaspectratio]{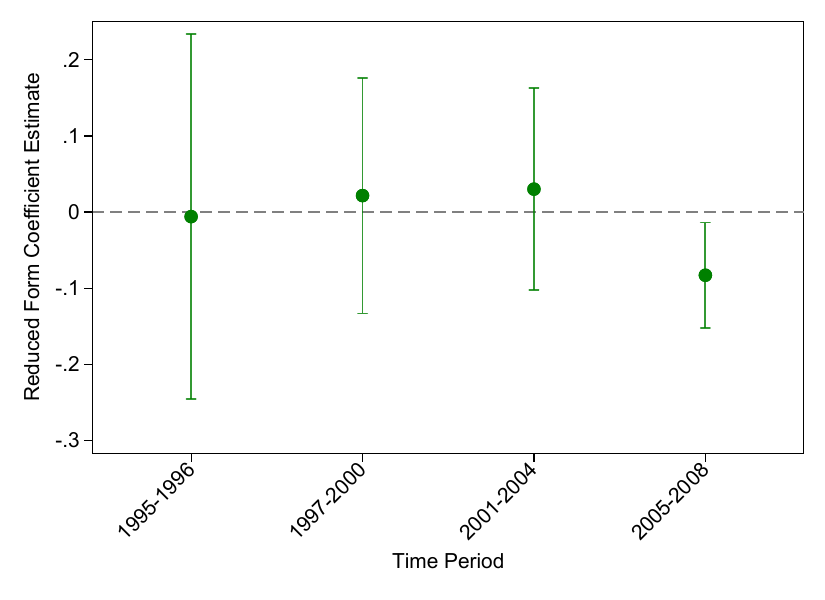}}
\vskip 0.1cm
\begin{minipage}{0.99\textwidth}
\vskip 0.1cm
{\footnotesize \emph{Notes:} Reduced form coefficients with newspapers as observation units. Each coefficient is from a separate regression mirroring the specification from Table \ref{tab:reduced_form}'s second column: the reduced form with state fixed effects, as well as demographic and channel controls and observations weighted by circulation. The outcome is Republican slant based on historical newspaper articles (see periods indicated on the horizontal axis), and the underlying models to quantify slant are based on models from historical Congress speeches. 
We obtain the newspaper articles from NewsLibrary only for fewer outlets in earlier periods: for 95-96, the coefficient is based on 139 newspapers, for 97-00 on 340 newspapers, and for 01-04 on 503 (relative to 718 in our main sample). The fourth estimate is our main observation period (05-08) but using the same sample as for 01-04 (i.e., 503 outlets). \par}
\end{minipage}
\end{figure}

Figure \ref{fig:timedyn} looks at timing. It shows the reduced form effect of FNC's channel position on newspaper slant for different periods.\footnote{We use the reduced form and not 2SLS because we only have viewership data for 2005-2008. While viewership varies substantially over time, the channel positions are relatively persistent \citep{MartinYurukoglu2017AER}.} Each coefficient comes from a separate regression using the reduced-form specification from Figure \ref{fig:reduced-form}: regressing slant on FNC channel position with state fixed effects, demographic, and channel controls. For each plotted coefficient, the outcome is the newspaper slant based on articles from the respective period (with periods indicated on the horizontal axis). To calculate these historical slant measures, we train separate models based on Congressional speeches from the respective period (see Appendix \ref{app:alt-model}). For example, for 1995-1996, we use a given newspaper's articles from that period and predict their slant using a model trained with Congressional speeches from that same period.\footnote{As for the main results, we do not train slant models year-by-year but pool several years for robust models.}

As the figure shows, the FNC channel position is not associated with newspaper slant before the channel's introduction. This is reassuring from an identification perspective, as it shows the channel position was not endogenously selected with prior newspaper slant. The wide confidence interval reflects the limited number of newspapers available in the dataset, primarily due to the lower coverage of earlier years in NewsLibrary. For instance, in 1995-1996, we could retrieve articles for only 139 newspapers out of the 718 in our main sample.\footnote{We ensure that non-findings for earlier periods are not due to different samples right below}.

Next, we take all the post-introduction years for which we have data (1997-2008). We take these twelve years and divide them equally into three periods: 1997-2000, 2001-2004, and 2005-2008. In 1997-2000, we have data for 340 outlets. We do not find any effect of FNC exposure on slant; the coefficient is of the opposite sign, close to zero (0.02), and insignificant. For 2001-2004, where we observe 503 newspapers, we find a similarly sized, again insignificant coefficient. Next, we look at our main observation period, 2005-2008, but only focus on the newspapers observed in 2001-2004. We still recover our effects (-0.08 with p$<$0.05).\footnote{The difference between 2001-2004 and 2005-2008 cannot be due to varying article numbers; the analysis builds on roughly 15 million articles for both periods, see Appendix \ref{sec:filter_no_articles}.}

First, from the very small and insignificant effect in the earlier years, we conclude that our main results are, indeed, driven by FNC and not pre-existing newspaper characteristics. Second, these results inform us on the potential mechanisms as our effects follow the channel's effect on Republican vote shares. \citet{DellaVignaKaplan2007TQJoE} and \citet{MartinYurukoglu2017AER} find that FNC had already increased the Republican vote share in 2000 by one-quarter to one-half of a percentage point.\footnote{This is the channel's effect at the extensive margin.} However, the effect grew considerably over time, reaching 6.34 percentage points in 2008 (\citeauthor{MartinYurukoglu2017AER}, p. 2568). Hence, our evidence is consistent with newspapers having shifted their slant over time in response to changed political preferences in their county community.\footnote{Appendix \ref{sec:app:timingcheck} confirms the robustness of Figure \ref{fig:timedyn}.} Our results align with \cite{gentzkow2010mediaslant}, who find that newspapers serving well-defined local markets respond to readers' political preferences in those markets. Hence, our study can be seen as a strong test of these findings by \citeauthor{gentzkow2010mediaslant}, since we move beyond the cross-sectional correlation between local audience preferences and slant and show that slant follows exogenous shifts in reader preferences.

We also look at geographical congruence in the effect on newspapers with the effect on voters in the cross section. Appendix \ref{sec:app:het} builds causal forest models to estimate conditional average treatment effects across counties, for both news slant and vote shares, following the method in \citet{athey2019generalized}. We show that counties with higher predicted treatment effects from our instrument on 2008 Republican vote shares tend to be counties where local newspapers shift more toward Republican language. This adds additional evidence in support of persuasion being an important explanation for our effects.

\paragraph{Cost-saving through borrowing of content.}

Next, we discuss cost-saving motives in terms of journalistic time use. Holding political preferences constant, pure cost-saving motives could manifest where producers use more readily available content and, in a potential trade-off between what readers want and cheaper content, the cost-saving outweighs the effect of losing some readers.\footnote{For instance, \citet{martin_mccrain_2019} discuss a similar trade-off regarding cheaper national news and potentially preferred local news in local TV stations.} Cost motives could play a role when right-wing content becomes more available. However, those motives \textit{alone} are unlikely to explain our effects -- for two reasons. First, if citizens change their political preferences (as the previous paragraph suggests), they plausibly change their local news reporting tastes -- unless local news reporting is fully decoupled from national political preferences. Such a full decoupling is unlikely in the light of evidence that cable TV has reshaped local policy preferences \citep{ash2023cable}.

The second reason cost motives are unlikely to fully explain our effects lies in the topics whose slant shifts. We can show that local newspapers weave the Republican slant into their original \textit{local} reporting.\footnote{Appendix \ref{sec:app:local_newspaper} describes how we distinguish local from non-local news.} As outlined in Appendix \ref{app:sec:local}, we recover our main results when we construct, for each newspaper, the slant measure only based on the articles covering local news. Most articles in our overall sample cover local news (73$\%$ on average at the newspaper level). This rules out verbatim copy-pasting of the predominantly national (or international) FNC \textit{stories} as a major driver of our results. Of course, the right-wing \textit{language} could still be woven into local stories simply because it is more available. However, given the size of our effect (just focusing on local news, we still find that if viewership increases by one standard deviation, articles move to the right by almost one-quarter of the difference between FNC and MSNBC) and the likely presence of persuasion, we see cost-saving motives as a possibly complementary but incomplete explanation for our results.

\paragraph{Market competition.} A final potential mechanism is that newspapers shift their ideological positioning due to market competition with Fox News. We do not have strong evidence either way on this mechanism. However, based on empirical and theoretical arguments outlined in Appendix \ref{sec:app:competition}, we conclude that market competition is not a major driver of our results.

\section{Conclusion}\label{sec:conc}

We document that exposure to partisan cable television, specifically Fox News Channel (FNC), influences the reporting of local newspapers. We quantify the slant of local news content by scaling it to speeches by Republican and Democrat speakers in Congress. We show that the local news slant becomes more Republican in counties with a higher quasi-experimental exposure to FNC. To this end, we use the channel's position in the cable lineup as an instrument for FNC viewership. Looking at the timing, we conclude that local newspapers likely change their slant in response to shifted policy priorities in their community (e.g., readers persuaded by the messaging of FNC). Our evidence also suggests the newspapers shift the slant in their local reporting. This means newspapers do not simply reproduce the stories from FNC (since the latter is nationally oriented), but weave the right-wing slant into their own, original reporting. 

These results add to the literature on the political effects of biased news. We provide new evidence on how partisan media influences not just voting and policies but also the content of other media organizations. Our results highlight that media outlets or technologies cannot necessarily be considered independent from each other. In addition, future work must allow for secondary indirect effects of partisan media. Such effects could work on voting and other outcomes by reshaping news landscapes and influencing the content of other media providers. 

Indeed, our research leaves as an open question whether the contagious-slant effect had further downstream effects, for example on the voting behaviour of newspaper readers. The observed shift in local news content might amplify FNC's effect in persuading voters to become more right-wing. Especially, since FNC increases right-wing slant on local-news topics -- which FNC as a national outlet does not cover -- that could induce rightward shifts in policy preferences on local issues that FNC messaging on its own would not. The evidence in \citet{ash2023cable} on FNC's effect on local fiscal policies could be an example of such secondary impacts. These findings highlight how national partisan media can shape local discourse even without direct links such as ownership, adding to concerns about the diminished public debate when local media's role as a moderating voice is weakened \citep[e.g.,][]{djourelova2021impact,darr2018newspaper}.

\clearpage\bibliographystyle{apalike}
\bibliography{new}

%% file: tables/main_instr.tex
\begin{table}[ht!]
\centering
\small
\begin{threeparttable}
\caption{2SLS: Newspaper Slant and Fox News Viewership \label{tab:main_instr}}
\begin{tabular}{l*{3}{r}} \hline
\noalign{\vspace{2ex}}
& \multicolumn{3}{c}{\textbf{Newspaper Slant}} \\ [1ex]

& (1) & (2) & (3) \\
\hline \hline

\input{tables/main_instr_raw} \vspace{-2.8ex} \\
\midrule 
State Fixed Effects   &         \checkmark     &  \checkmark &  \checkmark \\
Demographic Controls &         \checkmark   &         \checkmark  &         \checkmark \\
Channel Controls &            &         \checkmark   &         \checkmark \\
Newspaper Language Controls &            &           &         \checkmark \\
\bottomrule
\end{tabular}
\end{threeparttable}
\begin{minipage}{0.99\textwidth}
\vskip 0.2cm
{\footnotesize \emph{Notes:} Each column represents a separate regression. Observations are individual newspapers. All estimates are obtained using two-stage least squares (2SLS). The main explanatory variable is Fox News channel viewership instrumented with the channel's position (standardized). The outcome is the newspaper's Republican slant (standardized), measured as the share of Republican-leaning articles among all articles. All specifications include state fixed effects and demographic controls. Additionally, column (2) includes channel controls; column (3) further adds generic language controls. Appendix \ref{app:sec:sumstats} lists all control variables. We weight observations by circulation. We cluster standard errors at the state level (in parentheses; *** p<0.01, ** p<0.05, * p<0.1). \par}
\end{minipage}
\end{table}

%% file: tables/main_instr_raw.tex
\\ Fox News Channel Viewership&       0.718** &       0.493***&       0.515** \\
            &     (0.276)   &     (0.169)   &     (0.204)   \\

\\ \hline \multicolumn{1}{l}{Observations}&         718   &         718   &         718   \\
\multicolumn{1}{l}{K-P First-Stage F-Stat}&        9.01   &       14.70   &       15.06   \\

%% file: tables/main_proba.tex
\begin{table}[ht!]
\centering
\small
\begin{threeparttable}
\caption{\centering 2SLS: Newspaper Slant and Fox News Viewership\\
Robustness: Slant as Average of Raw Scores \label{tab:raw}}
\begin{tabular}{l*{3}{r}} \hline
\noalign{\vspace{2ex}}
& \multicolumn{3}{c}{\textbf{Republican}} \\
& \multicolumn{3}{c}{\textbf{Newspaper Slant}} \\ 
& \multicolumn{3}{c}{\textbf{Average of Score}} \\
[1ex]

& (1) & (2) & (3) \\
\hline \hline

\input{tables/main_proba_raw} \vspace{-2.8ex} \\
\midrule 
State Fixed Effects   &         \checkmark    &  \checkmark &  \checkmark \\
Demographic Controls &         \checkmark   &         \checkmark   &         \checkmark \\
Channel Controls &            &         \checkmark   &         \checkmark \\
Newspaper Language Controls &            &          &         \checkmark \\
\bottomrule
\end{tabular}
\end{threeparttable}
\begin{minipage}{0.99\textwidth}
\vskip 0.2cm
{\footnotesize \emph{Notes:} Each column represents a separate regression. Observations are individual newspapers. All estimates are obtained using two-stage least squares (2SLS). The main explanatory variable is Fox News channel viewership instrumented with the channel's position (standardized). This table is a robustness check, and the outcome is the newspaper's Republican slant (standardized) but measured as the average of the raw partisanship score of all articles (instead of the share of Republican-leaning articles as in Table \ref{tab:main_instr}). All specifications include state fixed effects and demographic controls. Additionally, column (2) includes channel controls; column (3) further adds generic language controls. Appendix \ref{app:sec:sumstats} lists all control variables. We weight observations by circulation. We cluster standard errors at the state level (in parentheses; *** p<0.01, ** p<0.05, * p<0.1). \par}
\end{minipage}
\end{table}

%% file: tables/main_proba_raw.tex
\\ Fox News Channel Viewership&       0.655*  &       0.310   &       0.414** \\
            &     (0.341)   &     (0.191)   &     (0.192)   \\

\\ \hline \multicolumn{1}{l}{Observations}&         718   &         718   &         718   \\
\multicolumn{1}{l}{K-P First-Stage F-Stat}&        9.01   &       14.70   &       15.06   \\

%% file: MS20230602-appendix-20251026.tex
\begin{appendices}

\setcounter{figure}{0} 
\renewcommand{\thefigure}{A.\arabic{figure}} 
\setcounter{table}{0} \renewcommand{\thetable}{A.\arabic{table}} 

\begin{center}
    {\huge \textbf{Online Appendices}}
\end{center} 

\section{Data Appendix} \label{sec:app:data}

\subsection{Newspaper Articles}
\label{app:news-data}

First, we provide some more information on NewsLibrary. For each article, the NewsLibrary provides the newspaper name, the headline, the date, the byline (if any), and (approximately) the first 80 words of the article. We focus on the first 80 words since, at the time of our data construction (June-August 2019), our subscription allowed us to access these article previews at large.\footnote{Full articles were available on a pay-per-piece basis, which was prohibitively expensive given our broad coverage in time and space.} An example of such a newspaper snippet is shown in Figure \ref{fig:newspap_example}.

\begin{figure}[ht!]
\caption{Example of a Local Newspaper Article Snippet} \label{fig:newspap_example}
\vskip 0.5cm
\input{tables/example_snippet}
\end{figure}

In principle, NewsLibrary encompasses almost 4,000 (to be precise, 3,966) unique outlets for 2005-2008. However, many outlets are not local newspapers, and they cannot be assigned to a county (e.g., the ``Army Communicator'' or the ``Air \& Space'' magazine). Furthermore, NewsLibrary often lists different editions of the same outlet separately. For instance, ``Augusta Chronicle, the (GA)'', ``Augusta Chronicle, the: Web Edition Articles (GA)'', and ``Augusta Chronicle, the: Blogs (GA)'' are listed separately. So, while our initial data collection for 2005-2008 covers 3,966 outlets (amounting to almost 44 million article snippets, see Section \ref{sec:filter_no_articles}), our main analyses focus on 718 outlets (and 15.1 million articles) that we can match to a county and for which county-level circulation data is available (see Section \ref{sec:data}). These 718 outlets are after collapsing different editions of the same outlet (as in the Augusta Chronicle example) to one observation because the ICPSR circulation data is typically not available separately for different editions of the same outlet.

For some outlets, not many articles are available. Section \ref{sec:app:slantmeasures} shows that our effects are robust to excluding outlets with an average of fewer than 1,000 articles per year (and, hence, potentially noisy slant measures).

\subsection{County Matching of Newspapers}
\label{app:sec:countymatching}

We match the sources listed in NewsLibrary to counties as follows: As a first step, we obtain the main county for each newspaper outlet based on the newspaper name and geographical information provided by NewsLibrary: For example, \textit{The Call (Woonsocket, RI)} or the \textit{Albany Democrat-Herald (OR)}. Where such information is unavailable or not conclusive, we also consult the U.S. Newspaper Directory and run manual web searches.
Then, as a second step, we match each newspaper to circulation data from the ICPSR. For this, we create a crosswalk using programmatic (exact and fuzzy) and manual matching. We verify each crosswalk entry manually. Using the crosswalk, we assign a newspaper's total circulation to the county we identified in the first step. We use 2004 total circulation for our main results (Section \ref{sec:main_results}) and also 1996 circulation in robustness checks or when exploring the dynamics (Sections \ref{sec:robust} and \ref{sec:mech}). 

\subsection{Filtering of the Article Snippets}
\label{sec:filter_no_articles}

Table \ref{tab:filter_no_articles} gives an overview of the number of articles collected and how we obtain the number of articles used in our main analyses and robustness checks.

\begin{table}[!htbp]
\centering
\caption{Number of Articles Collected and Filtering} \label{tab:filter_no_articles}
\input{tables/num_articles}
\end{table}

\clearpage

\subsection{Summary Statistics} \label{app:sec:sumstats}
\input{tables/summary_stat}

\clearpage

\section{Methods Appendix} \label{sec:app:methods}
\setcounter{figure}{0} 
\renewcommand{\thefigure}{B.\arabic{figure}} 
\setcounter{table}{0}
\renewcommand{\thetable}{B.\arabic{table}} 

This section presents additional material on measuring slant in newspaper articles using speeches from the Congressional Record.

\subsection{Text Pre-processing} \label{sec:app:preprocess}

We preprocess all texts (i.e., newspaper articles, congressional speeches) %
by converting the text to lowercase, removing punctuation and extra white spaces, expanding contractions, eliminating stopwords (including common stopwords like \textit{and} or \textit{or}, city and state names, and those specific to Congress and Congress members as defined by \citeauthor{gentzkow2019text}, \citeyear{gentzkow2019text}), as well as digits, timestamps, locations, and authorship metadata. Second, we perform tokenisation and stemming (employing the Porter stemming algorithm). Finally, we form bigrams (phrases of two words) and remove procedural bigrams, such as ``hear tell'' or ``time question,'' which are uninformative about the core topics of the text.

\subsection{Frequency Filtering} \label{sec:app:freqfilter}

Let $V_{k}$ be the bigram vocabulary used in speeches from members of party $k$, where $k \in \{\text{Republican}, \ \text{Democrat}\}$. We exclude speeches not given by Republicans or Democrats (e.g., Independents). Let $F_{k}^{b}$ be the frequency of bigram $b$ in speeches by members of party $k$. We construct $V_{\text{Rep}}$ and $V_{\text{Dem}}$. Then, we impose the following conditions on our vocabulary: a bigram must appear in at least 0.1\% of either Democrat or Republican speeches and 0.01\% of both Democrat and Republican speeches. The resulting set of bigrams is denoted as
\begin{equation*}
V = \left\{ b \in (V_{\text{Rep}}^{0.1} \cup V_{\text{Dem}}^{0.1}) \cap (V_{\text{Rep}}^{0.01} \cap V_{\text{Dem}}^{0.01}) \right\}
\end{equation*}

We remove infrequently used phrases to make computation faster (following, e.g., \citeauthor{gentzkow2019measuring}, \citeyear{gentzkow2019measuring}, or \citeauthor{cagelepennecmougin2024}, \citeyear{cagelepennecmougin2024}). The frequency threshold also makes it likely that the bigrams capture important political topics discussed in the media and not exclusively niche topics. Excluding bigrams used uniquely by one party ensures that the vocabulary reflects partisan language within the broader political discourse and not highly idiosyncratic terms -- this is again keeping in mind our goal of applying the model to newspaper language.
This procedure produces a vocabulary $V$ with 14,224 bigrams.

\subsection{Validation in the Congressional Record} \label{sec:app:validation}

To validate our approach, we use these predicted coefficients $\phi_b$ (see Section \ref{subsec:text-outcome}) to estimate partisanship for Republican and Democratic speeches. We determine the partisanship of a speech $n$ by projecting its observed relative bigram frequencies onto a partisan scale, formalized as:
\begin{equation}
\text{Speech\_Partisanship}_n = \sum_{b \in V} f_{bn} \cdot \phi_b
\end{equation}
where $\text{Speech\_Partisanship}_n$ represents the partisan score for speech $n$. %

Each speech's partisanship is, therefore, determined by the weighted sum of the bigram coefficients, with weights corresponding to bigram frequencies. Speeches with negative scores are classified as Democratic and those with positive scores as Republican. Our prediction accuracy is 0.64, and the corresponding confusion matrix is presented in Table \ref{tab:confusion-matrix}.\footnote{0.2\% of speeches are predicted with a score of 0. Given the binary nature of our sense-making exercise, we include them as Republican, which is inconsequential given the small share.} Looking at our slant measure directly instead of accuracy, Democrat speeches have an average score of 0.42 and Republican speeches of 0.71 -- meaning 42\% of Democrat speeches and 71\% of Republican speeches are predicted as Republican by the model -- with a difference of 0.29. Reassuringly, the model predicts higher scores for Republicans.

As an additional plausibility check, let us quantify the difference between FNC and MSNBC. The slant values are 0.48 and 0.34, respectively, giving a difference of 0.14. As expected, FNC is more Republican than MSNBC. The difference between the two channels is even larger when we only focus on the predicted partisan transcripts (around 83\%). Then, the difference between the channels is closer to the Republican-Democrat difference in Congress: 0.18 (0.58 vs. 0.40). For this check, we collected the cable news network transcripts for FNC and MSNBC prime-time shows from LexisNexis for the 2005-2008 period. We subject them to the same pre-processing steps as the Congressional speeches and then predict their partisanship.\footnote{For FNC, 12 shows were included (Fox News Edge, Fox News Sunday, Glenn Beck, Hannity, On the Record with Greta van Susteren, O'Reilly Factor, Special Report with Bret Baier, Special Report with Brit Hume, The Edge with Paula Zahn, The Five, The Kelly File, and Your World with Neil Cavuto), for MSNBC 18 shows (All In with Chris Hayes, Ashleigh Banfield On Location, Buchanan \& Press, Countdown with Keith Olbermann, Donahue, Hardball with Chris Matthews, Live with Dan Abrams, Morning Joe, Politics Nation, Race for the White House/1600 Pennsylvania Ave., Rita Cosby Live and Direct, Scarborough Country, The Ed Show, The Last Word with Lawrence O'Donnell, The News with Brian Williams, The Rachel Maddow Show, The Savage Nation, and Tucker), totaling 25,745 transcripts for the 2005-2008 period.}

\vskip 0.5cm

\begin{table}[ht!]
\centering
\caption{Prediction Performance for Identifying Congressional Speeches - GST Model 2005-2008} \label{tab:confusion-matrix}
\input{tables/congress_confusion_matrix_05_08}
\begin{minipage}{0.99\textwidth}
\vskip 0.2cm
{\footnotesize \emph{Notes:} Confusion matrix for congressional speeches (182,943 speeches). The top left cell represents true positives for the Democrat class; the bottom right cell represents true positives for the Republican class; the top right cell represents false negatives for Democrats; the bottom left cell represents false negatives for Republicans. The diagonal cells contain the true positive classifications for each class, and the off-diagonal cells contain the misclassifications. \par}
\end{minipage}
\end{table}

\subsection{Slant Models for Alternative Time Periods}\label{app:alt-model}

\input{tables/congress_speeches_per_period}

Following the methodology outlined in Section \ref{sec:text_data_methods:classify}, we estimate the model and bigram partisanship using the same vocabulary $V$, which was constructed over the entire 1995-2008 period, and apply it to the sub-periods 1995-1997, 1998-2000, 2001-2004, and 2005-2008. That is, we divided the entire period 1995-2008 into four (almost) equally sized periods to build four models for slant prediction. Table \ref{tab:congress_period} summarizes the number of congressional speeches. Table \ref{top-partisan-bigrams} presents the most partisan bigrams for Democrats and Republicans along with their estimated partisan coefficient by model. We also show each bigram's frequency in newspaper articles during the corresponding period.

\subsection{Most Partisan Phrases} \label{app:bigrams}

\begin{figure}[ht]
    \centering
     \caption{Wordcloud of top partisan bigrams over 2005-2008 - Fontsize represents frequency}
    \includegraphics[width=\textwidth]{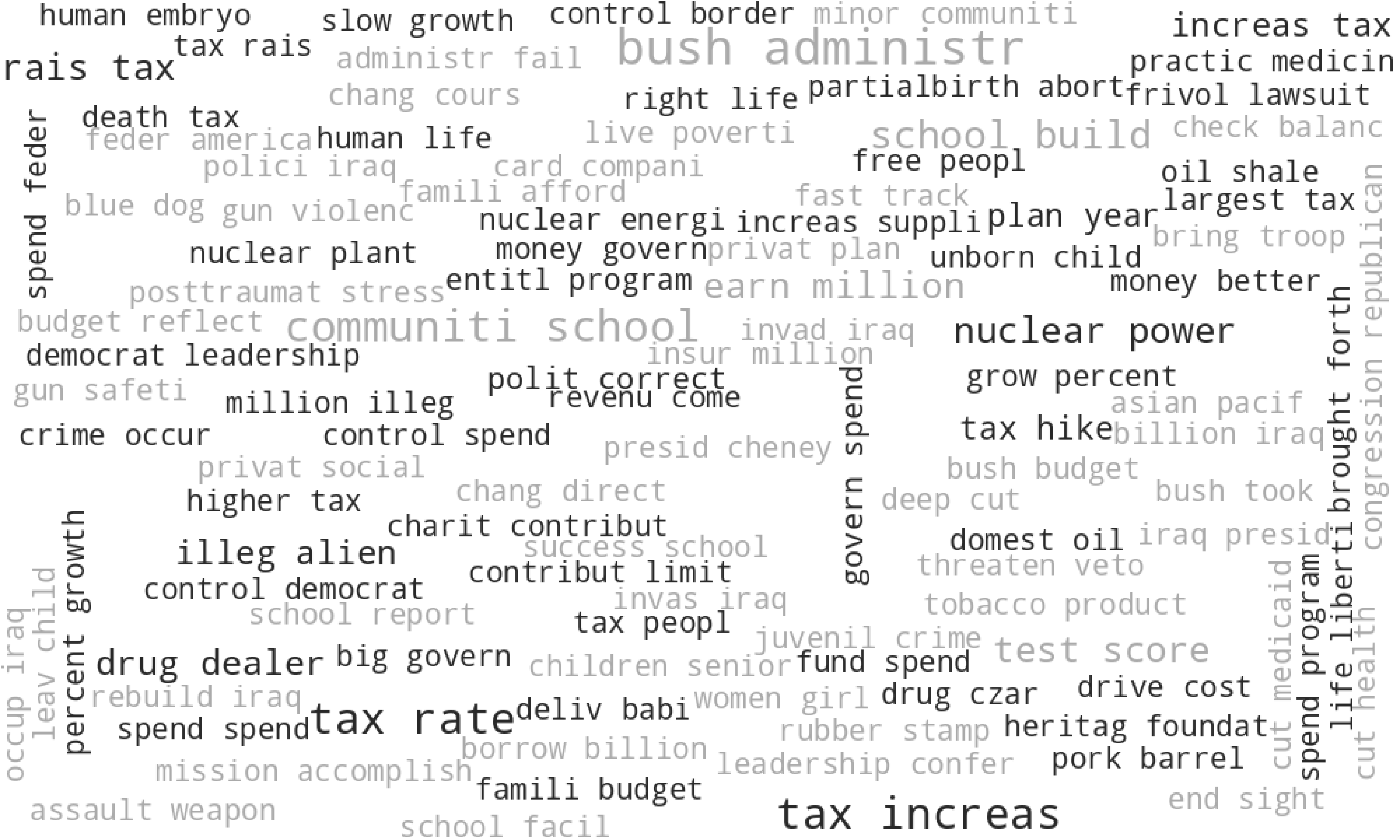}
    \label{fig:wordcloudfreq}
    \begin{minipage}{0.99\textwidth}
    \vspace{0.5cm}
    {\footnotesize \emph{Notes:} Most partisan bigrams for Democrats (light grey) and Republicans (dark grey) across newspapers during the period 2005--2008. We restrict the computation to the top 1,000 most partisan bigrams: font size represents the relative frequency of each bigram in the newspaper corpus, with larger text indicating higher occurrence. Procedural bigrams have been filtered out. \par}
\end{minipage}
\end{figure}

We provide additional visualizations of the most distinctive bigrams used to predict Republican and Democratic leanings in the context of newspaper snippets. Figure \ref{fig:wordcloudfreq} is similar to Figure \ref{fig:wordcloudcoef}: It also displays the most predictive bigrams for either Republican or Democratic speech in Congress. However, in \ref{fig:wordcloudfreq}, the font size represents the frequency of bigrams in newspapers. That is, it shows which partisan bigrams are frequently used in local newspaper articles.

\input{tables/partisan_bigrams_per_block}

\clearpage

\subsection{Distribution of Slant Across Newspapers}
\label{sec:app:distr:slant}

\begin{figure}[ht!]
\centering
\caption{Distribution of Slant Across Newspapers}
\includegraphics[width=0.7\textwidth]{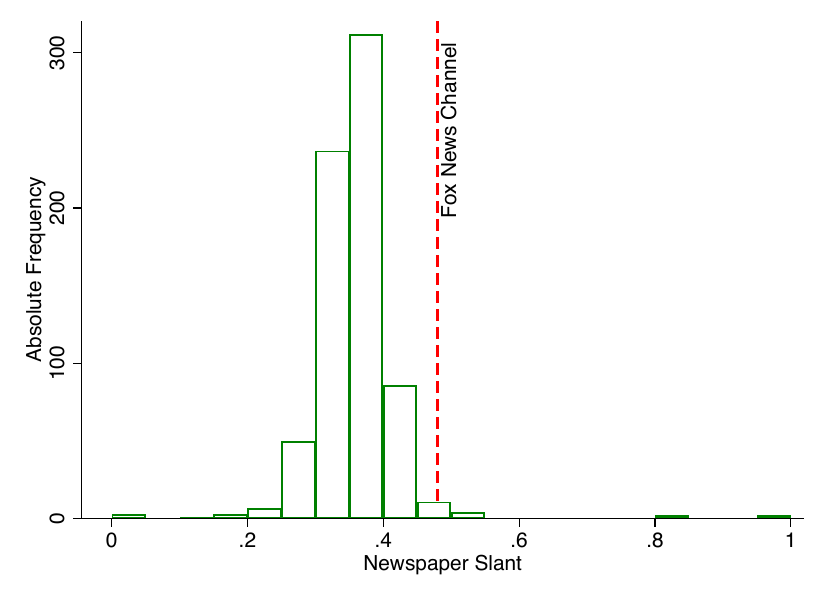}
\label{fig:hist_newspaper_predictions}
\begin{minipage}{0.99\textwidth}
{\footnotesize \emph{Notes:} Histogram of newspaper-level predictions. The figure shows the absolute frequency (unique newspaper counts) against the average, non-standardised estimated slant value for newspapers. The vertical line shows the average slant in FNC transcripts (at 0.48). \par}
\end{minipage}
\end{figure}

Figure \ref{fig:hist_newspaper_predictions} shows the distribution of the slant measure (unstandardised) for the 718 newspapers in our main sample.

\clearpage
\subsection{Example Articles by Partisan Slant} \label{sec:app:examplearticles}

\input{tables/top_rep_articles}

\input{tables/top_dem_articles}

\clearpage
\subsection{Removal of Routine Announcements} \label{sec:app:routine_announce}

Before aggregating the article slant at the newspaper level, we remove content related to routine announcements: births, deaths, obituaries, weddings, and notices of sale under power. While many of these announcements are correctly labelled as non-partisan, there are still false positives (e.g., through mentions of healthcare) that would make them false-positive regarding partisanship. At the newspaper level, the correlation is 0.95 when we compare the slant constructed with and without the routine announcements. In line with this, Appendix \ref{sec:app:slantmeasures} shows that our results are almost identical when not dropping those announcements.

We apply this filtering to both article titles and paragraph content using string matching. For titles, we exclude articles containing terms like ``death notice'' and ``obituary''. For paragraph content, we implement a more extensive filtering system that accounts for various formulations, including specific hospital records, yearly compilations (1997-2000), and funeral notices. Table \ref{tab:obifilter} presents the complete filtering criteria.

\begin{table}[htbp]
\centering
\caption{Routine Announcement Filtering Criteria}
\label{tab:obifilter}
\begin{tabular}{ll}
\hline
Field & Filter Criteria \\
\hline
\multirow{5}{*}{Title} & Contains: ``death notice'' \\
& Contains: ``obituaries'' \\
& Contains: ``obituary'' \\
& Contains: ``notice of sale under power'' \\
& Exact match: ``births'' \\
\hline
\multirow{19}{*}{Paragraph} & Starts with: ``births'' \\
& Starts with: ``deaths'' \\
& Starts with: ``rcal-births'' \\
& Starts with: ``local births'' \\
& Starts with: ``local deaths'' \\
& Starts with: ``all births below were recorded'' \\
& Starts with: ``all deaths below were recorded'' \\
& Starts with: ``these births were recorded at'' \\
& Contains: ``listing of births'' \\
& Contains: ``listing of deaths'' \\
& Contains: ``birth announcements'' \\
& Contains: ``announcements about births/deaths'' \\
& Contains: ``hospital notes'' \\
& Contains: ``[year] births'' \\
& Contains: ``funeral service'' \\
& Contains: ``funeral home'' \\
& Contains: ``michigan monroe county births'' \\
& Contains: ``hospital births'' \\
& Contains: ``wedding announcements'' \\
& Contains: ``special sections advertising links neighbors births'' \\
\hline
\end{tabular}
\end{table}

\clearpage
\subsection{Local Newspapers Classification} \label{sec:app:local_newspaper} 

To build a supervised classification model for predicting whether an article is local or non-local, we created a labelled dataset using GPT-3.5-Turbo with the following prompt: \textit{``You are given an article snippet, and your task is to determine if it is related to local news. Respond with 'local' or 'non-local'.''}

We randomly selected and labelled 100,000 newspaper articles published between 1995 and 2008. Afterward, the articles were cleaned using the methodology described in Section \ref{sec:app:preprocess}, this time forming unigrams (one-word phrases). We exclusively kept well-formatted GPT answers (either 'local' or 'non-local'), resulting in a dataset of 99,945 preprocessed and labelled articles, with 58.8$\%$ of them classified as local.

We split this labelled dataset into a training (80\%) and a test set (20\%) to train a logistic regression model on TF-IDF (Term Frequency-Inverse Document Frequency) unigrams extracted from the cleaned articles, with a maximum of 10,000 features selected based on term frequency. TF-IDF quantifies a word's significance by weighting its frequency in a specific newspaper snippet against its rarity across the entire newspaper training corpus. Our model's results on the 19,989 test articles -- 58.8$\%$ of which were labelled as local by GPT -- achieved an accuracy of 81$\%$, an F1 score of 84$\%$, precision of 83$\%$, and recall of 86$\%$. We validated our approach via blind human annotations of 2,108 newspaper article snippets, 72$\%$ of which were labelled as local by humans.\footnote{We recruited the annotators on Upwork. The annotators were instructed to read the newspaper article snippet and to decide whether it covered local or non-local news.} Our classifier achieves an accuracy of 77$\%$, an F1 score of 82$\%$, precision of 92$\%$, and recall of 75$\%$ on the human-labelled data. Table \ref{tab:top-local-features} shows the most significant unigrams for predicting local articles.
We use this well-validated model to label the articles from our main analysis. 60.8$\%$ articles are labelled as local.\footnote{In the summary statistics (Table \ref{tab:sumstat}), we show a slightly different share since that one is the average of the local share across newspapers.}

\input{tables/local_features_classification}

\clearpage

\section{Results Appendix}
\setcounter{figure}{0} 
\renewcommand{\thefigure}{C.\arabic{figure}} 
\setcounter{table}{0} 
\renewcommand{\thetable}{C.\arabic{table}} 

\subsection{First Stage Results} \label{sec:app:fs}

Appendix Table \ref{tab:first-stage} shows the first stage results, i.e., the association between the FNC channel position and FNC viewership (see the upper panel). The lower panel looks at the MSNBC position and MSNBC viewership.

In both panels, column (1) does not include controls, only census division fixed effects. Column (2) adds demographic controls (see Appendix Table \ref{tab:sumstat} for details). This leads the first stage to become statistically significant, and the Kleibergen-Paap cluster-robust first-stage F-statistic increases more than fourfold. In column (3), we add channel controls, further increasing the coefficient estimate and the F statistic for both panels.

Columns (4) to (6) mirror columns (1) to (3) but always include state instead of census fixed effects. The same pattern emerges: Both the coefficient estimates and especially the F statistics increase as we add the controls. Census division and state fixed effects result in similar estimates regarding size, significance, and F statistic. Therefore, we will concentrate on specifications along columns (5) and (6) for our main results and show the results for census fixed effects as a robustness check.

For FNC viewership, our preferred specification (column 6, with state fixed effects and all controls) implies that a one-standard-deviation increase in the FNC channel position decreases viewership by 19\% of a standard deviation. The Kleibergen-Paap cluster-robust first-stage F-statistic amounts to around 15, rendering our first stage relevant.\footnote{Notice that our F-statistic is smaller than Martin and Yurukoglu's (circa 15 in our preferred specification versus 39 in their preferred specification on p. 2576, Table 2, column 3). This fact aligns with us studying the relationship of the channel position and viewership at the county level. In contrast, their observations are at the zip code level; they, thus, draw on more variation (and a bigger sample).} The F statistic is small when using MSNBC as a predictor of FNC viewership; and it is also smaller with both FNC and MSNBC as instruments compared to FNC alone. We will thus use the FNC channel position as an instrument and the MSNBC position as a control (unless otherwise specified in robustness checks).

For the viewership of the MSNBC channel, the coefficient of its position amounts only to roughly -0.04 (while also being significant). Our paper largely focuses on the role of FNC in shifting newspaper slant (and only briefly discusses that MSNBC viewership does not have a robustly detectable effect on newspaper slant in Section \ref{sec:results:2sls}).

\input{tables/first_stage}

\clearpage
\subsection{Instrument Exogeneity} \label{sec:app:exogeneity}

This Section follows \cite{ash2023cable}. It demonstrates that our instrument is unrelated to demographic characteristics that predict news channel viewership or newspaper slant preferences. We use linear regressions with all our demographic and channel system characteristics (see Appendix Table \ref{tab:sumstat}) and state fixed effects as variables to predict viewership and newspaper slant. Specifically, we generate predictions for the endogenous regressor (viewership) and the outcome (the newspaper slant). These predictions capture the variation in viewership and news content due to pre-existing social, economic, and political county characteristics.

\input{tables/identification_check}

Appendix Table \ref{tab:identification-check} summarizes this identification check. We regress the predicted FNC viewership (column 1) and the predicted newspaper slant (column 2) on the position of FNC. None of the columns include demographic or channel controls (since those were used to predict the respective outcome). None of the columns show any significant relationship between the FNC position and the outcome. Overall, these results suggest that the channel positions capture variation that is not associated with county characteristics that are otherwise important for our endogenous regressor or outcome.

\clearpage
\subsection{OLS Results} \label{sec:app:ols}

Table \ref{tab:ols} shows OLS results for regressions of newspaper slant on FNC viewership.

The upper panel of the table looks at our main sample of newspapers. In column (1), we include no explanatory variables except FNC viewership (and state fixed effects). The viewership coefficient is marginally statistically significant (p$<$0.1). If the viewership of FNC increases by one standard deviation in a newspaper's county, its slant becomes more Republican by 0.13 standard deviations. This effect is smaller than our 2SLS estimate. While obvious confounders could lead to the 2SLS estimate being different from the OLS estimate (e.g., counties with pre-existing pro-Republican policy views tend to have both higher FNC viewership and more conservatively slanted newspapers), measurement error in the FNC viewership variable could lead to attenuation bias in the OLS estimates. The viewership coefficient becomes insignificant and smaller in the remaining three columns (2 to 4), where we introduce demographic, channel, and newspaper language controls. The coefficient size hovers around 0.07 to 0.10.  

The likely role of measurement error becomes apparent when examining a subsample of newspapers we use in robustness checks: those with more than 1,000 articles. For those outlets, we find larger and more precisely estimated effects: the OLS coefficient is 0.18 (significant at the 5\% level) without controls. It remains stable around 0.15 (and significant) after adding all controls. Focusing on this subsample likely leads to more precise slant estimates. Also, importantly, a newspaper's number of articles is correlated with the number of Nielsen respondents in its county (specifically, newspapers with $>$1000 articles tend to serve counties with respondent numbers that are 0.14 standard deviations higher). With higher respondent numbers, viewership is measured more precisely. Hence, our results are consistent with the instrumental variables addressing measurement error, while the OLS estimates may suffer from attenuation bias.\footnote{In line with this intuition, we obtain results highly similar to the lower panel of Table \ref{tab:ols} when we reduce the sample by dropping counties with the lowest Nielsen respondent numbers until we obtain a similar sample size (i.e., around 541). This similar sample size is obtained by dropping counties with Nielsen respondent numbers below the 25th percentile. Our main results are, if anything, stronger when using that sample in robustness checks (see Appendix Section \ref{sec:app:misc}).} Notably, this stark difference between the full sample and the $>$1000 articles sample appears only in OLS, not in our 2SLS estimates. Hence, overall, we interpret our OLS results as lending credence to our slant measure.

This interpretation is further supported by looking at the channel access variables. In both samples, the coefficients consistently align with the expected ideological leanings of the channels: the share of households with access to FNC shows a positive association with Republican slant, while MSNBC access shows a negative coefficient.

Even when focusing on newspapers with more than 1,000 articles, the OLS estimates remain substantially smaller than our 2SLS estimates (by a factor of two or three). This suggests that the larger 2SLS estimates likely reflect heterogeneous treatment effects: newspapers that change their slant in response to our instrument (the compliers) may be particularly susceptible to FNC influence. A similar argument about heterogeneous treatment effects among compliers is made by \cite{MartinYurukoglu2017AER} in their study of FNC effects on voting behavior. We discuss this argument in our context in more detail in Section \ref{sec:main_results}.

\vspace{1cm}

\input{tables/ols.tex}

\clearpage

\subsection{Reduced Form Results} \label{sec:app:rf}

Appendix Table \ref{tab:reduced_form} presents reduced-form results in tabular format. It shows the effects of the standardised FNC channel position on newspaper slant (also standardised). In column (1), we include state fixed effects and demographic controls. In column (2), we add channel controls. In column (3), we additionally include controls for generic newspaper language features (see Appendix Table \ref{tab:sumstat} for details on the different types of controls). The coefficient estimates are similar across the three columns and imply the following: When the FNC channel position increases by one standard deviation (which corresponds to around 11 positions), newspaper slant becomes \textit{less} Republican by roughly 0.09 standard deviations (notice that the coefficient sign is reversed compared to the 2SLS because we are looking at the position and not viewership here).

In column (2), we introduce the MSNBC position as a regressor along with the channel controls. The MSNBC position coefficient comes with a positive sign (suggesting that easier accessibility of MSNBC leads to more Democrat-leaning content). That effect is not robustly significant.

\input{tables/reduced_form.tex}

\clearpage

\subsection{MSNBC Results} \label{sec:app:msnbc}

Table \ref{tab:main_msnbc} replicates Table \ref{tab:main_instr} but focuses on instrumented MSNBC viewership instead of instrumented FNC viewership. As in Table \ref{tab:main_instr}, all columns include state fixed effects. The first column includes demographic controls. The second column also adds channel controls, and the third column further controls for generic newspaper language controls (see Appendix Table \ref{tab:sumstat} for details).

In all columns, the coefficients come with the expected negative sign -- higher MSNBC viewership implies less Republican slant. For example, in terms of magnitude, the third column suggests that newspaper slant is 0.42 standard deviations more Democrat if MSNBC viewership increases by one standard deviation. 

For MSNBC (relative to FNC in Table \ref{tab:main_instr}), the effect's size varies more across columns and are only weakly significant in column (2).

\vskip 1cm

\input{tables/main_msnbc}

\clearpage
\subsection{Robustness: Slant Measurement} \label{sec:app:slantmeasures}

Appendix Table \ref{tab:rep_over_partisan} mirrors Table \ref{tab:main_instr} but varies the outcome. In the main results, the outcome measures a newspaper's slant as the share of articles classified as Republican over the total of articles. Now, in Table \ref{tab:rep_over_partisan}, slant is the share of Republican articles over partisan articles: We divide the number of Republican articles by that of articles that are either Republican or Democrat. The results are almost identical to the main results, supporting the conclusion that our effects measure not merely a shift to Congress-like language (e.g., national politics) but, indeed, Republican over Democrat slant.

\input{tables/main_rep_over_partisan.tex}
Consistent with this, Table \ref{tab:demdum} shows no significant effects and negative coefficient signs when the outcome is Democrat slant (i.e., the number of Democrat articles over a newspaper's entire article count). These alternative measures are defined in Section \ref{subsec:text-outcome}.
\input{tables/main_demdum}

Next, Appendix Table \ref{tab:main_few1000} excludes outlets with less than a total of 1000 articles between 2005 and 2008 available on the NewsLibrary. This check rules out that our results are driven by observations with low article numbers, which might lead to spuriously extreme slant values (given the typical noisiness of text data measures). Finally, Table \ref{tab:main_keepobi} is based on newspaper slant measures where we do not drop routine announcements (e.g., birth and death announcements; see Section \ref{sec:app:routine_announce} for details). We still recover our results with similar coefficients.

\input{tables/main_few1000}

\input{tables/main_keepobi}

\clearpage

\subsection{Robustness: Circulation and Weighting} \label{sec:app:weights}

Table \ref{tab:main_weights_hist} replicates the main results from Table \ref{tab:main_instr} but weights each observation by the newspaper's circulation in the pre-FNC era (specifically, circulation in 1996). All coefficients remain qualitatively similar. Hence, our main results are not driven by potential circulation changes due to cable news exposure. Notice that we lose roughly 100 observations in this robustness check because, for these outlets, circulation data for 1996 is unavailable. 

\input{tables/main_weights_hist.tex}

Next, Table \ref{tab:main_weights_rel} shows robustness to using relative weights. That is, each newspaper's circulation is calculated as its market share in the county (based on our circulation data) times the county population. Hence, this exercise takes a county-level perspective: based on the county FNC viewership, how does the county-level newspaper landscape react? Again, the results remain similar; they become even slightly larger and more significant.

\input{tables/main_weights_rel.tex}

\clearpage

\subsection{Robustness: Fox News Relative to MSNBC Viewership} \label{sec:app:fncmsnbc}

Appendix Table \ref{tab:main_fncmsnbc} replicates Table \ref{tab:main_instr} but focuses on FNC viewership relative to MSNBC viewership. To this end, the endogenous variable is constructed as FNC viewership minus MSNBC viewership. Along these lines, the instrument is FNC channel position minus MSNBC channel position. We find effects similar to the main results, with slightly higher coefficient sizes.

\input{tables/main_fncmsnbc.tex}

\clearpage

\subsection{Robustness: Miscellaneous} \label{sec:app:misc}

Table \ref{tab:main_dropintab} demonstrates robustness to dropping counties with few Nielsen respondents. In these counties, the viewership is likely imprecisely measured; see discussion in Section \ref{sec:app:ols}. As we would expect, our effects become more precisely estimated when dropping these counties and, if anything, bigger.

\input{tables/main_dropintab.tex}

Table \ref{tab:main_censusdiv} shows that our results are robust to census division instead of state fixed effects. Table \ref{tab:main_instrnoclust} shows that our results remain robust when we use robust standard errors without clustering.

\begin{figure}%
\centering
\caption{\centering 2SLS: Newspaper Slant and Fox News Viewership \\
Robustness: Dropping States Individually \label{fig:leave_out_state}}
\includegraphics[width=0.5\linewidth]{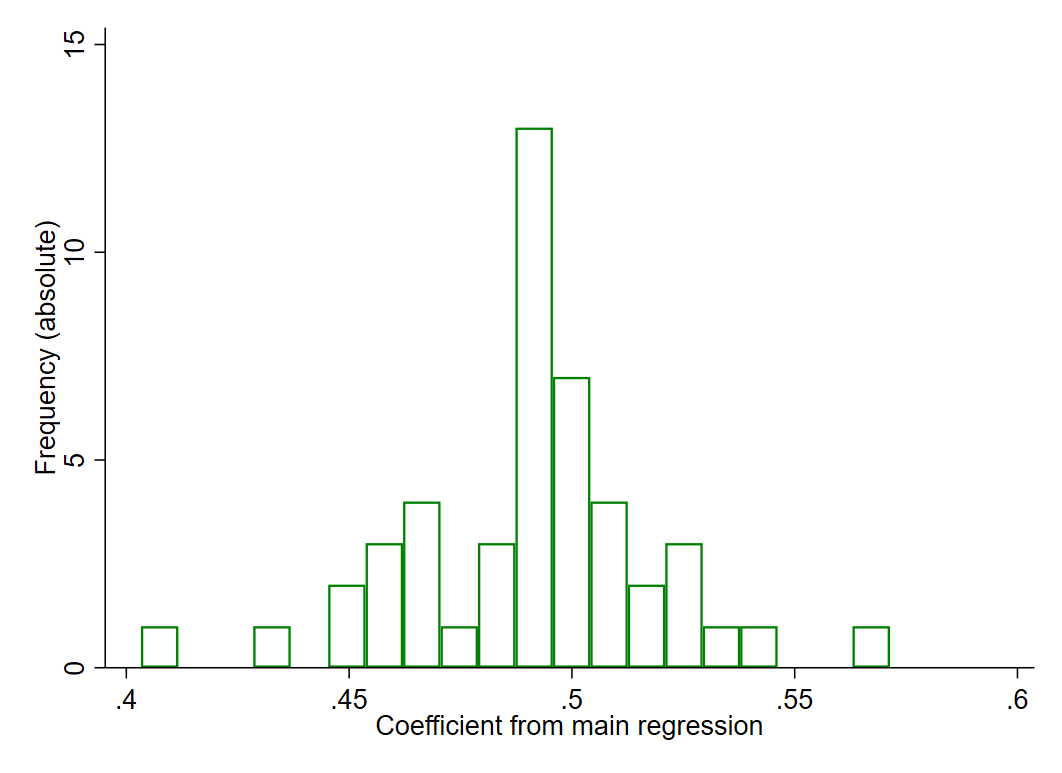}\includegraphics[width=0.5\linewidth]{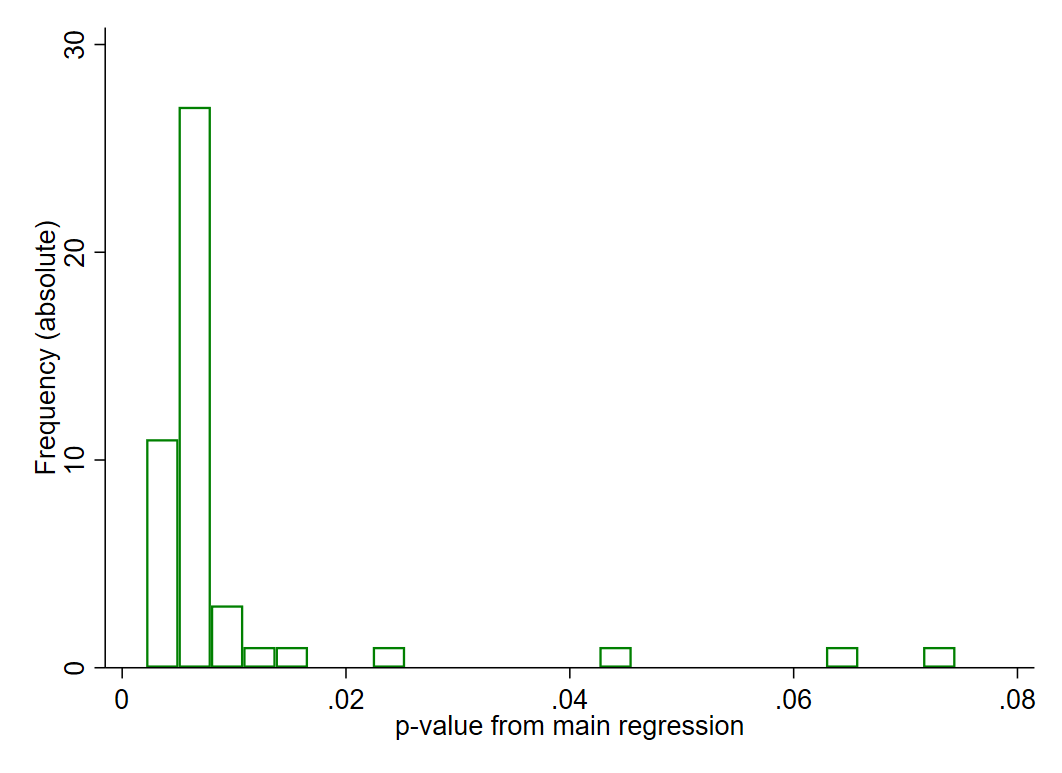}
\begin{minipage}{1\textwidth}
{\footnotesize \textit{Notes: }The histogram on the left shows FNC Viewership coefficients according to our preferred main specification (Table \ref{tab:main_instr}, column 3), but leaving out each state once. The histogram on the right shows the distribution of the p-values from the same regressions. \par}
\end{minipage}
\end{figure}

Furthermore, we conduct leave-one-out analyses at the state level. The results demonstrate stability: when excluding states (Figure \ref{fig:leave_out_state}) one at a time from our 2SLS specification (we use Table \ref{tab:main_instr}'s column 2), the estimated coefficients fall within a range of 0.40 to 0.60 (see left panel of the figure). These coefficients maintain statistical significance across iterations, as shown by the p-values concentrated near zero (see right panel). Our findings are not driven by any single state and, hence, by extension, an individual newspaper.

\vskip 1cm

\input{tables/main_censusdiv.tex}
\input{tables/main_instrnoclust}

\clearpage

\section{Mechanisms Appendix}

\renewcommand{\thefigure}{D.\arabic{figure}} 
\setcounter{table}{0} 
\setcounter{figure}{0}
\renewcommand{\thetable}{D.\arabic{table}} 

\subsection{Checks on Timing}
\label{sec:app:timingcheck}

\begin{figure}[ht!]
\centering
\caption{\centering Reduced Form: (Historical) Republican Newspaper Slant and Fox News Position \\ Robustness: Time-Varying Controls} \label{fig:timedyn2}
{\includegraphics[width=0.99\textwidth, keepaspectratio]{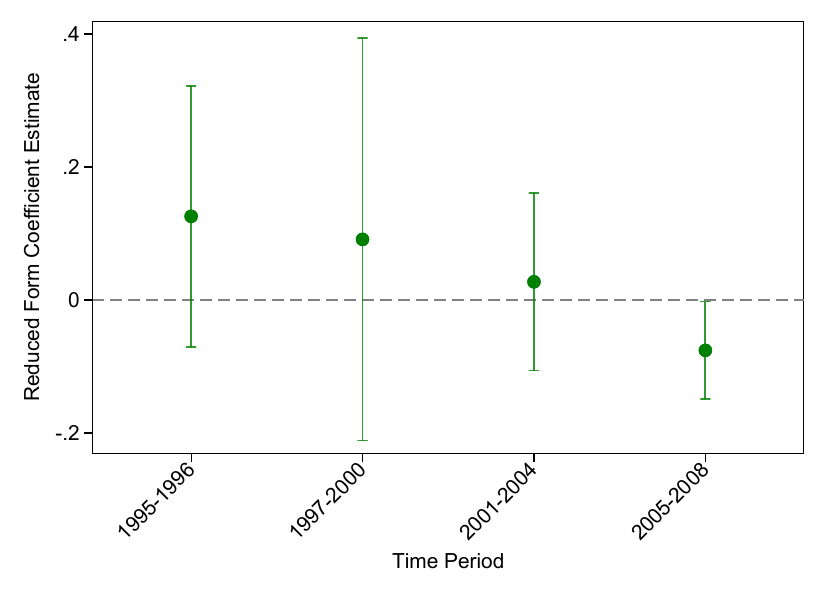}}
\vskip 0.1cm
\begin{minipage}{0.99\textwidth}
\vskip 0.1cm
{\footnotesize \emph{Notes:} Reduced form coefficients with newspapers as observation units. Each coefficient is from a separate regression mirroring the specification from Table \ref{tab:reduced_form}'s second column: the reduced form with state fixed effects, as well as demographic and channel controls and observations weighted by circulation. We use time-varying covariates (either from the 2000 or 2010 census) as a robustness check to Figure \ref{fig:timedyn}. The outcome is Republican slant based on historical newspaper articles (see periods indicated on the horizontal axis), and the underlying models to quantify slant are based on models from historical Congress speeches. 
We obtain the newspaper articles from NewsLibrary only for fewer outlets in earlier period: for 95-96, the coefficient is based on 139 newspapers, for 97-00 on 340 newspapers, and for 01-04 on 503 (relative to 718 in our main sample). The fourth estimate is our main observation period (05-08) but using the same sample as for 01-04 (i.e., 503) outlets. \par}
\end{minipage}
\end{figure}

Figure \ref{fig:timedyn} in Section \ref{sec:mech} shows that slant changes only become detectable in 2005-2008. Figure \ref{fig:timedyn2} replicates that figure but lets the covariates vary with the period. In the main part, we use census data from 2010. Now, for the earlier periods, we use the data from the 2000 census. We restrict the set of controls to variables available in both periods -- this explains why the coefficient for 2005-2008 is also slightly different from that in Figure \ref{fig:timedyn}. We find qualitatively equivalent results in both figures.

\begin{figure}[ht!]
\centering
\caption{\centering Reduced Form: (Historical) Democrat Newspaper Slant and Fox News Position} \label{fig:timedyn_demdum}
{\includegraphics[width=0.99\textwidth, keepaspectratio]{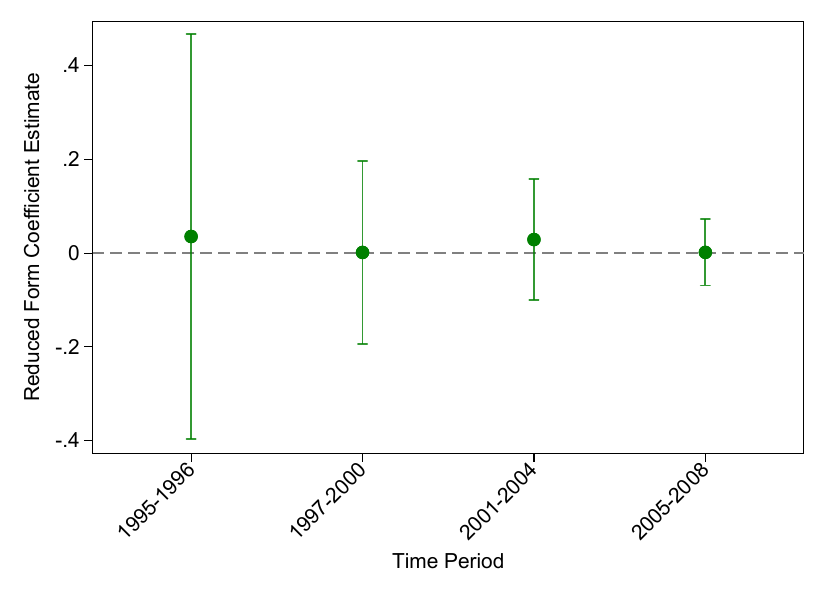}}
\vskip 0.1cm
\begin{minipage}{0.99\textwidth}
\vskip 0.1cm
{\footnotesize \emph{Notes:} Reduced form coefficients with newspapers as observation units. Each coefficient is from a separate regression mirroring the specification from Table \ref{tab:reduced_form}'s second column: the reduced form with state fixed effects, as well as demographic and channel controls and observations weighted by circulation. Contrary to Figure \ref{fig:timedyn}, the outcome is Democrat instead of Republican slant. The slant is based on historical newspaper articles (see periods indicated on the horizontal axis), and the underlying models to quantify slant are based on models from historical Congress speeches. 
We obtain the newspaper articles from NewsLibrary only for fewer outlets in earlier period: for 95-96, the coefficient is based on 139 newspapers, for 97-00 on 340 newspapers, and for 01-04 on 503 (relative to 718 in our main sample). The fourth estimate is our main observation period (05-08) but using the same sample as for 01-04 (i.e., 503) outlets. \par}
\end{minipage}
\end{figure}

Figure \ref{fig:timedyn_demdum} also replicates Figure \ref{fig:timedyn} but with the share of Democrat-leaning (instead of Republican-leaning) articles as the outcome. We do not find an effect in any period.

\clearpage
\subsection{Treatment Effect Heterogeneity}
\label{sec:app:het}

This section documents a geographical congruence: counties where the newspapers' slant becomes more Republican in response to higher FNC exposure also tend to be counties where the Republican vote shares increase.

Using the method from \citeauthor{athey2019generalized}'s \citeyear{athey2019generalized}, we predict, based on county-level covariates, which counties likely see large treatment effects from the FNC instrument on (i) their voting in 2008 and (ii) their newspapers' slant in 2005-2008. This method is a machine-learning approach -- a causal forest -- to model heterogeneous treatment effects across treated units in response to an instrument. It allows the reduced form effect of channel position to vary flexibly with local covariates. %
We train two models: The first one for newspaper slant as the outcome, and the second one with the 2008 Republican vote share as the outcome. For this exercise, we drop counties with low numbers of Nielsen respondents (below the 25th percentile) for better precision.

\begin{figure}[ht!]
\centering
\caption{Reduced Form: Predicted Treatment Effects on Slant and Republican Votes in 2008} \label{fig:het_predictions_binscatter}
\includegraphics[width=0.99\textwidth]{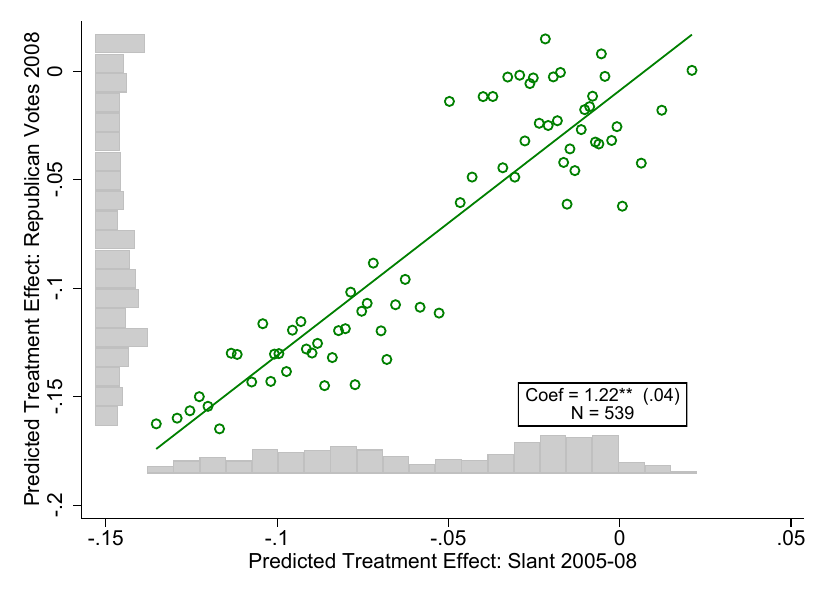}
\begin{minipage}{1\textwidth}
{\footnotesize \textit{Notes:} 
 Binned scatterplot of the standardised predicted treatment effect of the FNC channel position on Republican vote shares in 2008 (vertical axis) against the standardised predicted treatment effects of the same instrument on newspaper slant (horizontal axis). Cross-section with newspaper-level observations weighted by newspaper circulation. We include state fixed effects and channel controls (see Table \ref{tab:sumstat} for the specific controls) to improve the precision of the treatment effect estimation. The demographics are used to predict the treatment effect size. Next to the axes, we show the distributions of the underlying variables in light grey. \par}
\end{minipage}
\end{figure}

Using the trained causal forests, we predict the treatment effects for both outcomes. Figure  \ref{fig:het_predictions_binscatter} shows that the predicted effect sizes for each outcome are highly correlated across counties. The Pearson's correlation is 0.77. Table \ref{tab:most_assoc_covs} shows the covariates most associated with high treatment effects.

Overall, these findings suggest that the types of counties that are persuaded by FNC regarding voting also have newspapers that are responsive in their slant. We interpret these descriptive results as evidence for the importance of persuasion, arguably in particular on the demand side.

\begin{table}[ht!]
\centering
\caption{Covariates Most Associated with High Treatment Effects (County-Level)} \label{tab:most_assoc_covs}
\input{tables/top_covariates}
\begin{minipage}{0.99\textwidth}
\vskip 0.2cm
{\footnotesize \emph{Notes:} Covariates most associated with a strong response of slant in 2005-2008 (left column) and Republican vote shares in 2008 (right column) to the FNC channel position. The covariates were identified using heterogeneous treatment effect estimation via instrumental variables as proposed by \cite{athey2019generalized}. The top-listed covariate represents the most associated one; the second covariate is the second-most associated one, etc. %
\par}
\end{minipage}
\end{table}

\clearpage

\subsection{Slant in Articles on Local News} \label{app:sec:local}

As Section \ref{sec:mech} mentioned, our main results hold when we construct, for each newspaper, the slant measure only with articles covering local news, as Table \ref{tab:mech_local} illustrates. Column (1) replicates Table \ref{tab:main_instr}'s column (3) but with the share of local articles as the outcome. Columns (2) to (4) replicate our main specifications (columns 1 to 3 of Table \ref{tab:main_instr}), but with newspaper slant only based on articles covering local news. Appendix \ref{sec:app:local_newspaper} details how we distinguish local from higher-level reporting. The coefficient is smaller than in the main results, but the differences are statistically insignificant. Hence, even for local news, we find sizeable effects.\footnote{When looking at non-local articles, we find an effect of similar magnitude (0.42). It is insignificant, likely due to low observation numbers (around 1600 non-local articles per newspaper on average).}

\vskip 1cm

\input{tables/mech_local.tex}

\clearpage

\subsection{Analyses on Market Competition} \label{sec:app:competition}

This appendix provides some suggestive evidence on whether market competition is an important driver of our results. First, we look at circulation. Table \ref{tab:reduced_form_circ} replicates Table \ref{tab:reduced_form} but with circulation as the outcome and without circulation weights. Table \ref{tab:reduced_form_circ_hist} repeats this exercise but additionally controls for circulation in 1996. We do not see any effects of the FNC channel position on circulation. If counties with lower FNC channel positions are under higher competitive pressures than counties with higher positions, we may expect this to manifest in lower circulation.

Second, as shown in Appendix \ref{app:sec:local}, we find similar slant effects when focusing on local reporting only (i.e., focusing on articles covering local as opposed to national or international news). While local and national news outlets can be competitors (e.g., over advertising revenues as shown by \citeauthor{Angelucci2024}, \citeyear{Angelucci2024}), there is little direct competition from FNC over the content of local news. We hypothesize that the effect would be different if market competition were an important driver of our results.

Finally, from a theoretical viewpoint, conventional approaches of spatial competition, e.g., those in the spirit of Hotelling (1929), would rather predict the left-wing shift of the positions of existing outlets in response to the entry of a right-wing competitor due to differentiation incentives. Consider, for example, a model by \cite{EatonLipsey1975}: multiple outlets compete exclusively over locations along [0, 1] line with uniformly distributed voters. If there were, say, four outlets before entry, in the unique equilibrium pattern, two outlets position themselves at 1/4 and two at 3/4. If a right-wing competitor enters at 5/6, for instance, in a new equilibrium, there will be two outlets and 1/6, one outlet at 1/2, and the fourth outlet will join the entrant at 5/6. That is, most of the outlets would move to the left in response to the right-wing entrant. We find the opposite.

Overall, we interpret this evidence and argumentation as going against market competition being a pivotal mechanism in our setting.

\input{tables/reduced_form_circ}

\input{tables/reduced_form_circ_controlhist}

\end{appendices}

%% file: tables/example_snippet.tex
\begin{center}
\begin{minipage}[t]{0.6\textwidth}
\textit{\textbf{Aberdeen American News}\\[1ex]}
\textit{WET CONDITIONS LEAVE RIPE CROPS EXPOSED TO THREATS, INCLUDING HUNGRY ANIMALS LAND LINES\\[1ex]}
\textit{October 22, 2007}
\end{minipage}
\vskip 0.5cm
\begin{minipage}[t]{0.6\textwidth}
\textit{Harvest is behind schedule in northeast South Dakota and in the state as a whole because of wet weather last week.A year ago Saturday, the soybean harvest was 93 percent completed in South Dakota, according to the Farm Forum. This year, it was 61 percent finished in the state before the wet spell began Oct. 13, but little has been harvested since, according to the National Agricultural Statistics Service.A year ago, the Farm Forum reported that frosty conditions during the week [...]}
\end{minipage}
\end{center}

%% file: tables/num_articles.tex
\begin{tabular}{l|c|c}
\toprule
\textbf{Filtering} & \textbf{\# Articles} & \textbf{Results} \\ 
\hline \\
\multicolumn{3}{l}{\textit{\textbf{All periods:}}} \\
\cmidrule{1-3}
Raw scrapes 1995-2008  & 93,731,393 & None (impossible: no \\  & & county assignment) \\
\hline \\
\multicolumn{3}{l}{\textit{\textbf{Main sample 2005-2008:}}} \\
\cmidrule{1-3}
Raw scrapes & 43,495,343 & None (impossible: no \\  & & county assignment) \\
\cmidrule{1-3}
Thereof matched to circulation data & 16,006,097 &  None (impossible: no \\
 & & county assignment) \\
\cmidrule{1-3}
Thereof in regression (matched to county) & 15,116,215 &  Table \ref{tab:main_instr}  \\
\bottomrule \\
\multicolumn{3}{l}{\textit{\textbf{Historical sample 2001-2004:}}} \\
\cmidrule{1-3}
Raw scrapes & 25,768,302 & None (impossible) \\
\cmidrule{1-3}
Thereof matched to circulation data & 17,317,353 & None (impossible) \\
\cmidrule{1-3}
Thereof in regression (matched to county) & 16,021,408 &  Figure \ref{fig:timedyn}  \\
\bottomrule \\
\multicolumn{3}{l}{\textit{\textbf{Historical sample 1997-2000:}}} \\
\cmidrule{1-3}
Raw scrapes & 17,714,873 & None (impossible) \\
\cmidrule{1-3}
Thereof matched to circulation data & 11,934,736 & None (impossible) \\
\cmidrule{1-3}
Thereof in regression (matched to county) & 11,110,070 &  Figure \ref{fig:timedyn}  \\
\bottomrule \\
\multicolumn{3}{l}{\textit{\textbf{Placebo sample 1995-1996:}}} \\
\cmidrule{1-3}
Raw scrapes & 6,752,875 & None (impossible) \\
\cmidrule{1-3}
Thereof matched to circulation data & 4,609,973 & None (impossible) \\
\cmidrule{1-3}
Thereof in regression (matched to county) & 4,307,634 &  Figure \ref{fig:timedyn}  \\
\bottomrule
\end{tabular}

%% file: tables/summary_stat.tex
\begin{table}[htbp]\centering
\caption{Summary Statistics \label{tab:sumstat}}
\footnotesize
\begin{tabular}{p{8.2cm}p{1cm}p{1cm}p{1cm}p{1cm}p{0.9cm}}
\toprule
Variable & Mean & STD & Min & Max & N \\ 
\midrule
\textit{\textbf{Newspapers}} & & & & \\ \hline

\input{tables/summary_stat_raw_1} 
\textit{\textbf{News Channels}} & & & & \\ \hline

\input{tables/summary_stat_raw_2}

\end{tabular}
\end{table}

\begin{table}[htbp]\centering
\caption*{Table \ref{tab:sumstat} Continued: Summary Statistics}
\footnotesize
\begin{tabular}{p{8.2cm}p{1cm}p{1cm}p{1cm}p{1cm}p{0.9cm}}
\toprule
Variable & Mean & STD & Min & Max & N \\ 
\midrule
\textit{\textbf{Demographics}} & & & & \\ \hline

\input{tables/summary_stat_raw_3}
\end{tabular}
\end{table}

%% file: tables/summary_stat_raw_1.tex
05-08 main slant measure (share Rep. articles)&        0.36&        0.07&        0.00&        1.00&         718\\
05-08 alternative slant (as main but incl. obituaries)&        0.35&        0.07&        0.00&        1.00&         718\\
05-08 alternative slant (avg. article-level partisanship)&        0.00&        0.03&       -0.39&        0.36&         718\\
05-08 main slant only for local news&        0.35&        0.07&        0.00&        1.00&         718\\
05-08 Democrat slant (share Dem. articles)&        0.33&        0.06&        0.00&        1.00&         718\\
05-08 alternative slant (share Rep. among partisan)&        0.52&        0.06&        0.00&        1.00&         717\\
01-04 main slant (share Rep. articles)&        0.21&        0.06&        0.00&        1.00&         504\\
01-04 Democrat slant (share Dem. articles)&        0.49&        0.07&        0.00&        1.00&         504\\
97-02 main slant (share Rep. articles)&        0.18&        0.09&        0.00&        1.00&         342\\
97-02 Democrat slant (share Dem. articles)&        0.52&        0.13&        0.00&        1.00&         342\\
95-96 main slant (share Rep. articles)&        0.19&        0.12&        0.00&        1.00&         151\\
95-96 Democrat slant (share Dem. articles)&        0.53&        0.14&        0.00&        1.00&         151\\
Share of local news &        0.73&        0.12&        0.24&        1.00&         718\\
Circulation (1000s) &       40.88&       74.64&        0.75&      652.43&         718\\
Circulation 96 (1000s)&       45.17&       80.16&        0.92&      634.63&         608\\
Word length         &        4.75&        0.09&        4.38&        5.51&         718\\
Article length      &       85.70&        3.54&       64.28&      102.00&         718\\
Vocabulary size     &        0.11&        0.11&        0.02&        0.74&         718\\
Collected articles/year (1000s)&       21.05&       33.30&        0.00&      309.15&         718\\

 \hline

%% file: tables/summary_stat_raw_2.tex
FNC channel position 2005-2008&       42.30&       11.07&        5.00&       71.83&         718\\
MSNBC channel position 2005-2008&       45.13&       12.47&       13.25&      108.25&         718\\
Ratings \% FNC 2005-2008&        0.55&        0.28&        0.00&        3.48&         718\\
Ratings \% MSNBC 2005-2008&        0.15&        0.49&        0.00&       13.00&         718\\
Share pop. access to CNN 2005-2008&        0.97&        0.06&        0.05&        1.00&         718\\
Share pop. access to FNC 2005-2008&        0.96&        0.07&        0.04&        1.00&         718\\
Share pop. access to MSNBC 2005-2008&        0.93&        0.13&        0.03&        1.00&         718\\

Number of Nielsen survey respondents&         604&         960&           3&        7052&         718\\

 \bottomrule \bottomrule

%% file: tables/summary_stat_raw_3.tex
Population (1000s)  &         411&        1048&           6&        9819&         718\\
Population density  &         675&        2410&           3&       33886&         718\\

Republican vote share 1996&        0.43&        0.10&        0.14&        0.68&         718\\
Share Whites        &        0.79&        0.16&        0.21&        0.99&         718\\
Share Blacks        &        0.10&        0.12&        0.00&        0.77&         718\\
Share Asians        &        0.03&        0.05&        0.00&        0.44&         718\\
Share Hispanics     &        0.12&        0.15&        0.00&        0.96&         718\\
Share males         &        0.49&        0.01&        0.45&        0.61&         718\\
Share age 10-29     &        0.41&        0.05&        0.24&        0.64&         718\\
Share age 30-59     &        0.40&        0.03&        0.27&        0.49&         718\\
Share urban         &        0.47&        0.41&        0.00&        1.00&         718\\
Share high school graduates&        0.31&        0.07&        0.12&        0.53&         718\\
Share some college education&        0.29&        0.05&        0.15&        0.41&         718\\
Share Bachelor's degree&        0.16&        0.06&        0.05&        0.41&         718\\
Share postgraduate degree&        0.09&        0.04&        0.03&        0.27&         718\\
Share food stamp recipients&        0.12&        0.05&        0.01&        0.32&         718\\
Share management occupations&        0.32&        0.06&        0.19&        0.56&         718\\
Share service occupations&        0.18&        0.03&        0.11&        0.32&         718\\
Share office occupations&        0.24&        0.02&        0.17&        0.31&         718\\
Share construction occupations&        0.03&        0.01&        0.01&        0.12&         718\\
Share production occupations&        0.14&        0.05&        0.03&        0.32&         718\\
Share arts occupations&        0.03&        0.01&        0.01&        0.10&         718\\
Share same-sex households&        0.00&        0.00&        0.00&        0.02&         718\\
Share homeowners    &        0.69&        0.08&        0.22&        0.86&         718\\
Median tax rate     &        0.01&        0.01&        0.00&        0.03&         718\\
Aggregated taxes (million)&       13.12&       10.52&        0.41&       55.56&         718\\
Median household income (1000s)&       50.54&       12.28&       24.10&       95.11&         718\\
Median housing values (1000s)&      184.43&      126.65&       53.29&      826.73&         718\\

 \bottomrule \bottomrule

%% file: tables/congress_confusion_matrix_05_08.tex
\begin{tabular}{lcc}
\toprule 
 & \textbf{Predicted Democrat} & \textbf{Predicted Republican}\\
\midrule 
\textbf{Actual Democrat} & 31.58\% (58K) & 23.25\% (43K) \\
\textbf{Actual Republican} & 12.92\% (24K) & 32.26\% (59K)\\
\bottomrule
\end{tabular}

%% file: tables/congress_speeches_per_period.tex
\begin{table}[ht!]
\centering
\caption{Number of congress speeches per model} \label{tab:congress_period}
\begin{tabular}{lcccc}
\toprule 
Model Period & \textbf{2005-2008} & \textbf{2001-2004} & \textbf{1998-2000} & \textbf{1995-1997} \\
\midrule
Number of Congress Speeches & 182,943 & 171,852 & 143,747 & 171,637 \\
Share of Republican Speeches (\%) & 45.2 & 46.7 & 55.1 & 55.4 \\ 
\bottomrule
\end{tabular}
\end{table}

%% file: tables/partisan_bigrams_per_block.tex
\begin{center}
\begin{footnotesize}
\begin{longtable}{@{\hskip 0.5cm} c | l r r | l r r @{\hskip 0.5cm}}
\caption{Most partisan bigrams by period model \label{top-partisan-bigrams}} \\
 \\
\toprule
\textbf{Period} & \textbf{Democrat} & \textbf{Coef} & \textbf{Count} & \textbf{Republican} & \textbf{Coef} & \textbf{Count} \\
\midrule
\endfirsthead

\toprule
\textbf{Period} & \textbf{Democrat} & \textbf{Coef} & \textbf{Count} & \textbf{Republican} & \textbf{Coef} & \textbf{Count} \\
\midrule
\endhead

\bottomrule
\endfoot

2005--2008 & gun safeti & -3.55 & 579 & liabil reform & 3.51 & 105 \\
2005--2008 & surplus trillion & -2.69 & 27 & margin tax & 3.05 & 125 \\
2005--2008 & gun violenc & -2.51 & 1806 & unborn child & 2.92 & 1709 \\
2005--2008 & break wealthi & -2.50 & 101 & crime introduc & 2.89 & 18 \\
2005--2008 & redeploy troop & -2.48 & 197 & size govern & 2.82 & 261 \\
2005--2008 & occup iraq & -2.45 & 1006 & medic liabil & 2.78 & 322 \\
2005--2008 & trillion surplus & -2.44 & 37 & unborn children & 2.77 & 350 \\
2005--2008 & dog coalit & -2.39 & 84 & largest tax & 2.69 & 496 \\
2005--2008 & assault weapon & -2.37 & 1954 & control spend & 2.65 & 548 \\
2005--2008 & privat social & -2.33 & 1596 & deliv babi & 2.53 & 1388 \\
2005--2008 & wealthiest american & -2.28 & 251 & repeal death & 2.51 & 162 \\
2005--2008 & republican refus & -2.22 & 259 & harm come & 2.50 & 120 \\
2005--2008 & tobacco product & -2.18 & 1878 & entitl spend & 2.49 & 104 \\
2005--2008 & asian pacif & -2.14 & 596 & death tax & 2.48 & 1408 \\
2005--2008 & lowincom work & -2.13 & 252 & tax hike & 2.46 & 4735 \\
2005--2008 & cut wealthi & -2.13 & 295 & partialbirth abort & 2.41 & 734 \\
2005--2008 & presid cheney & -2.11 & 1808 & growth feder & 2.34 & 121 \\
2005--2008 & hispan caucus & -2.07 & 168 & marriag tax & 2.22 & 40 \\
2005--2008 & administr fail & -2.05 & 672 & twothird favor & 2.20 & 14 \\
2005--2008 & invad iraq & -2.04 & 2196 & growth spend & 2.19 & 188 \\
2005--2008 & leav child & -1.98 & 593 & duti defend & 2.16 & 52 \\
2005--2008 & school facil & -1.95 & 3298 & human embryo & 2.16 & 1125 \\
2005--2008 & school build & -1.93 & 12932 & unfund liabil & 2.11 & 375 \\
2005--2008 & deficit debt & -1.89 & 34 & elimin marriag & 2.09 & 21 \\
2005--2008 & cut wealthiest & -1.88 & 82 & miguel estrada & 2.04 & 14 \\
2005--2008 & corpor profit & -1.88 & 1099 & law send & 2.04 & 205 \\
2005--2008 & earn million & -1.88 & 9090 & percent growth & 2.04 & 1816 \\
2005--2008 & fail polici & -1.85 & 561 & million illeg & 2.02 & 3055 \\
2005--2008 & republican join & -1.85 & 891 & polit correct & 2.02 & 4186 \\
2005--2008 & unfortun republican & -1.84 & 47 & rais tax & 2.00 & 9488 \\
\midrule
2001--2004 & unfortun republican & -2.97 & 24 & harm come & 3.52 & 117 \\
2001--2004 & republican refus & -2.80 & 201 & crime introduc & 3.37 & 21 \\
2001--2004 & wealthiest percent & -2.64 & 198 & duti defend & 3.18 & 62 \\
2001--2004 & republican allow & -2.61 & 110 & peopl illeg & 2.86 & 438 \\
2001--2004 & congression caucus & -2.51 & 474 & repeal death & 2.63 & 76 \\
2001--2004 & work longer & -2.47 & 548 & liabil reform & 2.60 & 185 \\
2001--2004 & break wealthi & -2.44 & 89 & medic liabil & 2.57 & 707 \\
2001--2004 & congression republican & -2.43 & 1273 & societi like & 2.51 & 321 \\
2001--2004 & wealthiest american & -2.36 & 217 & death tax & 2.46 & 1190 \\
2001--2004 & dog coalit & -2.35 & 33 & terribl crime & 2.43 & 162 \\
2001--2004 & budget fail & -2.31 & 244 & enforc enhanc & 2.40 & 17 \\
2001--2004 & asian pacif & -2.23 & 530 & budget estim & 2.39 & 431 \\
2001--2004 & defeat previous & -2.21 & 320 & law send & 2.38 & 161 \\
2001--2004 & smaller class & -2.19 & 721 & rule control & 2.38 & 128 \\
2001--2004 & administr refus & -2.16 & 436 & defend citizen & 2.36 & 74 \\
2001--2004 & republican leadership & -2.15 & 1283 & deliv babi & 2.27 & 1363 \\
2001--2004 & deep cut & -2.12 & 1343 & growth spend & 2.22 & 120 \\
2001--2004 & republican friend & -2.12 & 169 & control propon & 2.22 & 10 \\
2001--2004 & peopl color & -2.10 & 936 & control rank & 2.21 & 11 \\
2001--2004 & republican propos & -2.09 & 659 & appropri budget & 2.19 & 200 \\
2001--2004 & republican budget & -2.08 & 321 & chang heart & 2.17 & 1612 \\
2001--2004 & cut republican & -2.08 & 211 & rule natur & 2.10 & 98 \\
2001--2004 & massiv tax & -2.08 & 357 & sought recognit & 2.03 & 15 \\
2001--2004 & deficit debt & -2.05 & 44 & exceed hour & 2.00 & 63 \\
2001--2004 & decent wage & -2.04 & 158 & tribe tribal & 1.96 & 73 \\
2001--2004 & republican continu & -2.03 & 263 & unfund liabil & 1.96 & 263 \\
2001--2004 & billion short & -2.02 & 147 & suspend rule & 1.94 & 249 \\
2001--2004 & blue dog & -2.01 & 368 & taken twothird & 1.94 & 14 \\
2001--2004 & privat social & -2.00 & 880 & like describ & 1.93 & 225 \\
2001--2004 & red ink & -1.99 & 2217 & illeg alien & 1.93 & 2610 \\
\midrule
1998--2000 & unfortun republican & -3.49 & 19 & follow prayer & 4.01 & 92 \\
1998--2000 & cut wealthiest & -3.29 & 38 & room russel & 3.51 & 25 \\
1998--2000 & republican friend & -3.27 & 85 & ago feder & 2.97 & 580 \\
1998--2000 & wealthiest percent & -3.14 & 123 & feder cost & 2.93 & 39 \\
1998--2000 & leadership refus & -3.01 & 32 & increas trillion & 2.70 & 12 \\
1998--2000 & gun safeti & -2.92 & 1180 & unborn children & 2.66 & 197 \\
1998--2000 & break wealthi & -2.92 & 23 & heritag area & 2.66 & 119 \\
1998--2000 & republican refus & -2.88 & 126 & rule control & 2.62 & 78 \\
1998--2000 & hispan caucus & -2.81 & 80 & receiv testimoni & 2.55 & 11 \\
1998--2000 & wealthiest peopl & -2.76 & 73 & yesterday feder & 2.54 & 1680 \\
1998--2000 & republican leadership & -2.76 & 1134 & death tax & 2.38 & 951 \\
1998--2000 & summer job & -2.65 & 1299 & cost estim & 2.33 & 1982 \\
1998--2000 & congression caucus & -2.58 & 329 & transport meet & 2.24 & 240 \\
1998--2000 & decent wage & -2.57 & 80 & repeal death & 2.22 & 119 \\
1998--2000 & gun violenc & -2.49 & 1708 & tribe tribal & 2.21 & 39 \\
1998--2000 & bush republican & -2.45 & 1090 & growth govern & 2.20 & 99 \\
1998--2000 & wealthiest american & -2.39 & 54 & dirksen offic & 2.18 & 29 \\
1998--2000 & reproduct health & -2.38 & 239 & communist china & 2.11 & 378 \\
1998--2000 & friend republican & -2.38 & 65 & exceed hour & 2.08 & 45 \\
1998--2000 & discharg petit & -2.36 & 55 & defens submit & 2.04 & 19 \\
1998--2000 & peopl color & -2.35 & 656 & congression design & 2.03 & 10 \\
1998--2000 & smaller class & -2.32 & 868 & employe offic & 2.02 & 534 \\
1998--2000 & econom educ & -2.28 & 380 & action debat & 1.98 & 42 \\
1998--2000 & card compani & -2.27 & 1128 & minut control & 1.97 & 117 \\
1998--2000 & year bush & -2.23 & 277 & public agenc & 1.94 & 1090 \\
1998--2000 & deep cut & -2.23 & 352 & passeng motor & 1.93 & 18 \\
1998--2000 & defeat previous & -2.23 & 195 & hardearn money & 1.91 & 274 \\
1998--2000 & urg republican & -2.22 & 196 & militari depart & 1.90 & 58 \\
1998--2000 & labor right & -2.22 & 139 & feder bureaucrat & 1.88 & 122 \\
1998--2000 & cut wealthi & -2.19 & 45 & minut second & 1.86 & 14492 \\
\midrule
1995--1997 & decent wage & -3.18 & 80 & follow prayer & 5.18 & 58 \\
1995--1997 & republican friend & -3.12 & 57 & room russel & 3.56 & 24 \\
1995--1997 & republican refus & -3.03 & 137 & judiciari meet & 2.85 & 48 \\
1995--1997 & wealthiest american & -2.94 & 91 & marriag tax & 2.55 & 41 \\
1995--1997 & break wealthi & -2.93 & 67 & growth govern & 2.54 & 106 \\
1995--1997 & cost prescript & -2.83 & 55 & percent electr & 2.29 & 159 \\
1995--1997 & gun violenc & -2.80 & 407 & ago feder & 2.21 & 471 \\
1995--1997 & wealthi american & -2.75 & 87 & transport meet & 2.16 & 174 \\
1995--1997 & wealthiest percent & -2.70 & 30 & receiv testimoni & 2.13 & 24 \\
1995--1997 & children defens & -2.69 & 430 & yesterday feder & 2.10 & 2081 \\
1995--1997 & cut wealthiest & -2.67 & 18 & death tax & 2.07 & 400 \\
1995--1997 & unfortun republican & -2.66 & 19 & communist china & 2.06 & 354 \\
1995--1997 & democrat altern & -2.48 & 86 & unborn child & 2.05 & 526 \\
1995--1997 & star war & -2.43 & 2847 & nativ claim & 2.03 & 13 \\
1995--1997 & congression caucus & -2.43 & 313 & rule control & 2.01 & 82 \\
1995--1997 & summer job & -2.42 & 1602 & passeng motor & 1.98 & 18 \\
1995--1997 & live wage & -2.37 & 334 & congression design & 1.97 & 26 \\
1995--1997 & gun law & -2.37 & 518 & hardearn money & 1.95 & 207 \\
1995--1997 & question defeat & -2.32 & 20 & execut agenc & 1.91 & 117 \\
1995--1997 & deep cut & -2.31 & 701 & dirksen offic & 1.90 & 28 \\
1995--1997 & pay prescript & -2.31 & 50 & tax freedom & 1.88 & 141 \\
1995--1997 & republican leadership & -2.25 & 1168 & largest tax & 1.87 & 260 \\
1995--1997 & cut educ & -2.22 & 343 & margin tax & 1.81 & 49 \\
1995--1997 & center budget & -2.22 & 259 & suspend rule & 1.80 & 112 \\
1995--1997 & medicar cut & -2.19 & 217 & salari expens & 1.78 & 111 \\
1995--1997 & educ cut & -2.18 & 186 & elimin marriag & 1.77 & 17 \\
1995--1997 & medicaid cut & -2.16 & 131 & emerg requir & 1.77 & 33 \\
1995--1997 & wage worker & -2.11 & 207 & partialbirth abort & 1.76 & 1689 \\
1995--1997 & peopl color & -2.10 & 460 & inform pleas & 1.75 & 299 \\
1995--1997 & tax break & -2.10 & 6327 & present law & 1.74 & 153 \\
\end{longtable}
\begin{minipage}{0.99\textwidth}
\vskip 0.1cm
{\footnotesize \setstretch{1} \emph{Notes:} Top 30 most partisan bigrams for Democrats and Republicans, along with their estimated partisan coefficients and their frequency across newspaper articles during the period, based on each model per period. Procedural bigrams were filtered out. \par}
\end{minipage}
\end{footnotesize}
\end{center}

%% file: tables/top_rep_articles.tex
\begin{table}[ht]
\fontsize{9.5}{10.5}\selectfont
\caption{Newspaper articles with the strongest Republican partisanship \label{tab:news-sim-rep}}    
\textbf{Daily News (Bowling Green, KY), March 15, 2006} \\ 
\textbf{Partisanship coefficient: 2.96} \\ 
\textit{Letters} \\ 
Special Sections Advertising Links Letters Medical liability reform never had a chance Your article March 4, "Medical liability bill fails,"; was unsurprising.Expecting real medical liability tort reform from a legislature controlled by lawyers is ludicrous. Also predictable were local lawyers' defenses of our current lawyers' paradise.Exceptional cases (like penis removal) are presented, with suggestions that lawyers only cost (greedy) insurance
\medskip

\textbf{Lincoln Journal Star: Web Edition Articles (NE), January 19, 2005} \\ 
\textbf{Partisanship coefficient: 2.92} \\ 
\textit{Bill highlights fetal pain issue} \\ 
Women who seek late-term abortions would automatically get information telling them that the unborn child can feel pain and asking if they want the fetus to have anesthesia during the abortion process, under a bill sponsored by Lincoln Sen. Mike Foley.Foley said LB752 is based on a growing body of evidence that an unborn child of 20 weeks, and perhaps younger, can feel pain. Session '05 The bill is modeled after the federal "fetal pain awareness act" that has been introduced in
\medskip

\textbf{Chattanooga Times Free Press (TN), March 18, 2005} \\ 
\textbf{Partisanship coefficient: 2.50} \\ 
\textit{What about the lives it saves?} \\ 
The United Nations, ever on the lookout for ways to condemn Israel, has come up with a new plan of attack. It already has condemned the security wall Israel is building to keep out Palestinian suicide bombers, and the U.N. World Court has said Israel has to tear down the wall - regardless of the nation's justified concerns about security.Now, U.N. Secretary-General Kofi Annan says the United Nations will keep a careful accounting of all the supposed harm that comes to
\medskip

\textbf{The Wichita Eagle (KS), September 2, 2005} \\ 
\textbf{Partisanship coefficient: 2.46} \\ 
\textit{BLOGATORIAL: EXCERPTS FROM OUR WEB LOG} \\ 
Allergic to tax hikes - and proud of it It will surprise no one who follows the Statehouse that the Wichita-area delegation includes some of the Kansas Taxpayers Network's best-loved legislat ors. In the Wichita-based group's 2005 legislative ratings, perfect 100 percent scores were accorded Republican Sens. Phil Journey of Haysville and Peggy Palmer of Augusta. The House standouts, with ratings of 105, included Reps. Bonnie Huy of Wichita, Dick Kelsey of Goddard, Steve Huebert of Valley
\medskip

\textbf{The Deseret News (Salt Lake City, UT), February 13, 2005} \\ 
\textbf{Partisanship coefficient: 1.90} \\ 
\textit{No licenses for illegals} \\ 
I just finished reading the headlines in the Deseret Morning News (Feb. 10) and could not believe my eyes. "Senators target license abuses," with a sub-head "Utah leaders seeking a new type of driving privilege for illegal aliens."Why in the world are our illustrious legislators seeking to create any privilege for people who are illegally in this country? It would seem to me they should be seeking solutions and means for identifying them and then deporting them. From the numbers
\medskip

\centering
\begin{minipage}{0.99\textwidth}
{\footnotesize \setstretch{1} \emph{Notes:} Examples among the top 10,000 Republican-leaning articles for the period 2005--2008. The partisanship coefficient here refers to article-level partisanship. \par}
\end{minipage}

\end{table}

%% file: tables/top_dem_articles.tex
\begin{table}[ht]
\fontsize{9.5}{10.5}\selectfont
\caption{Newspaper articles with the strongest Democrat partisanship \label{tab:news-sim-rep}}    
\textbf{The Roanoke Times (VA), February 24, 2006} \\ 
\textbf{Partisanship coefficient: -3.55} \\ 
\textit{DONT GAG DOCTORS ABOUT GUN SAFETY} \\ 
Doctors should make a deal with the General Assembly: Physicians won't attempt to legislate if legislators quit trying to practice medicine.A bill that would gag pediatricians who discuss gun safety with the parents of their patients passed the Virginia House of Delegates last week, and is under consideration in the Senate.Gun ownership is just one issue pediatricians bring up with parents when going over safety checklists during their children's routine checkups.That
\medskip

\textbf{The Kansas City Star (MO), December 4, 2005} \\ 
\textbf{Partisanship coefficient: -2.48} \\ 
\textit{Democrats leaning toward strategic redeployment} \\ 
PHOENIX - Democrats are starting to unify behind a get-out-of-Iraq strategy that would urge the redeployment of U.S. troops to other countries, party chairman Howard Dean said Saturday. "The Democrats now have vision around 
\medskip

\textbf{The St. Augustine Record (FL), August 16, 2006} \\ 
\textbf{Partisanship coefficient: -2.45} \\ 
\textit{Letter: Welcome to Wonderland} \\ 
Editor: Be afraid. Be very afraid.The lunatics have taken over the asylum and they've stolen all the guns.Israel's mass slaughter of Lebanese civilians, a minor mirror image of Bush's "untidy" occupation of Iraq, is only the latest installment of what the right-wing extremists believe is going to be our future, if they can help it, of "unending war." After all, somebody has to keep the weapons industry productive. It is one of the few businesses we haven't moved out of the
\medskip

\textbf{Lincoln Journal Star: Web Edition Articles (NE), February 7, 2005} \\ 
\textbf{Partisanship coefficient: -2.33} \\ 
\textit{Protesters carry on without foes} \\ 
AFL-CIO members on Tuesday were ready to vent their frustrations at Edward Jones. Dozens gathered outside the brokerage house's North Ridge Shopping Center office with signs reading "Hands off my Social Security."Ken Mass, president of the Nebraska AFL-CIO, said the protesters, mostly blue-collar workers from across the state, came to show displeasure at what it describes as Edward D. Jones \& Co.'s support for the privatization of Social Security.But the office was closed -
\medskip

\textbf{The Oklahoman (Oklahoma City, OK), February 3, 2008} \\ 
\textbf{Partisanship coefficient: -2.28} \\ 
\textit{Author says taxpayers foot bill for wealthy Americans riches} \\ 
We have always heard there is no such thing as a free lunch, and the author agrees with this with one exception: the very wealthy who benefit from government subsidies and handouts paid for by the taxpayers.In "Free Lunch: How the Wealthiest Americans Enrich Themselves at Government Expense, and Stick You With the Bill" (Portfolio, \$24.95), David Cay Johnston writes that government benefits and subsidies are available only to corporations and those people rich enough to own a large
\medskip

\centering
\begin{minipage}{0.99\textwidth}
{\footnotesize \setstretch{1} \emph{Notes:} Examples among the top 10,000 Democrat-leaning articles for the period 2005--2008. The partisanship coefficient here refers to article-level partisanship. \par}
\end{minipage}

\end{table}

%% file: tables/local_features_classification.tex
\begin{table}[ht]
\small
\caption{Most Predictive Unigrams for Local Classification\label{tab:top-local-features}}    
\centering
\begin{tabular}{lr|lr}
\hline
\textbf{Local} & \textbf{Coef} & \textbf{Non-local} & \textbf{Coef} \\
\hline
local & 10.71 & dear & -4.45 \\
counti & 10.03 & fy & -3.93 \\
polic & 5.87 & american & -3.77 \\
area & 5.84 & bush & -3.63 \\
communiti & 5.42 & china & -3.59 \\
gov & 5.18 & ny & -3.58 \\
downtown & 5.14 & french & -3.54 \\
citi & 5.12 & calif & -3.44 \\
resid & 4.88 & british & -3.38 \\
council & 4.31 & york & -3.35 \\
mayor & 4.13 & germani & -3.23 \\
town & 4.06 & movi & -3.18 \\
superintend & 3.85 & marri & -3.16 \\
student & 3.66 & global & -3.14 \\
park & 3.59 & investor & -3.11 \\
school & 3.56 & european & -3.04 \\
said & 3.56 & world & -3.01 \\
herald & 3.39 & nation & -2.98 \\
north & 3.23 & engag & -2.92 \\
center & 3.16 & nyse & -2.90 \\
airport & 3.15 & type & -2.85 \\
township & 3.13 & date & -2.81 \\
street & 3.06 & network & -2.78 \\
librari & 3.02 & nfl & -2.77 \\
staff & 3.01 & nba & -2.66 \\
locat & 3.00 & america & -2.65 \\
roanok & 2.99 & ep & -2.61 \\
region & 2.89 & fla & -2.60 \\
chamber & 2.88 & billion & -2.58 \\
event & 2.83 & washington & -2.56 \\
\hline
\end{tabular}
\begin{minipage}{0.99\textwidth}
\vskip 0.2cm
{\footnotesize \setstretch{1}  \emph{Notes:} Top 30 most predictive unigrams for classifying newspaper articles as 'Local' or 'Non-local,' along with their estimated predictive coefficients.}
\end{minipage}
\end{table}

%% file: tables/first_stage.tex
\begin{table}[ht!]
\centering
\resizebox{\textwidth}{!}{%
\begin{threeparttable}
\caption{First Stage: Cable TV Position Effects on Viewership \label{tab:first-stage}}
\begin{tabular}{l*{6}{r}} \hline
\noalign{\vspace{2ex}}
& \multicolumn{6}{c}{\textbf{Fox News Viewership}} \\ [1ex]
& (1) & (2) & (3) & (4) & (5) & (6) \\
\hline \hline

\input{tables/first_stage_raw1} \vspace{-2.8ex} \\
\midrule \\[1ex]

 & \multicolumn{6}{c}{\textbf{MSNBC Viewership}} \\ [1ex]
& (1) & (2) & (3) & (4) & (5) & (6) \\
\hline \hline

\input{tables/first_stage_raw2} \vspace{-2.8ex} \\ 
\midrule 
Census Div. Fixed Effects &         \checkmark   &         \checkmark   &         \checkmark   &           &  &  \\
State Fixed Effects &            &            &            &         \checkmark     &  \checkmark &  \checkmark \\
Demographic Controls &            &         \checkmark   &         \checkmark   &             & \checkmark  & \checkmark \\
Channel Controls &            &            &         \checkmark   &              &   & \checkmark \\
\bottomrule
\end{tabular}
\end{threeparttable}
}
\begin{minipage}{0.99\textwidth}
\vskip 0.2cm
{\footnotesize \emph{Notes:} Each column represents a separate regression. The upper panel shows regressions of Fox News Channel viewership on its channel position, while the lower panel shows regressions of MSNBC viewership on its channel position. Both viewership and position variables are standardized. All observations are weighted by circulation. Columns (1)-(3) include census division fixed effects, while columns (4)-(6) include state fixed effects. Demographic controls and channel controls are introduced progressively (in columns 2 and 3 or 5 and 6, respectively). The Kleibergen-Paap cluster-robust first-stage F-statistics are shown for the Fox News position, the MSNBC position, and both jointly. All controls are listed in Appendix \ref{app:sec:sumstats}. We cluster standard errors at the state level (in parentheses; *** p<0.01, ** p<0.05, * p<0.1). \par}
\end{minipage}
\end{table}

%% file: tables/first_stage_raw1.tex
\\ Fox News Channel Position&      -0.098   &      -0.142***&      -0.173***&      -0.104   &      -0.152***&      -0.185***\\
            &     (0.068)   &     (0.046)   &     (0.044)   &     (0.081)   &     (0.051)   &     (0.048)   \\
[1ex] MSNBC Channel Position&               &               &       0.043   &               &               &       0.040   \\
            &               &               &     (0.026)   &               &               &     (0.027)   \\

\\ \hline \multicolumn{1}{l}{Observations}&         718   &         718   &         718   &         718   &         718   &         718   \\
\multicolumn{1}{l}{K-P F-Stat FNC}&        2.09   &        9.51   &       15.61   &        1.63   &        9.01   &       14.70   \\
\multicolumn{1}{l}{K-P F-Stat MSNBC}&               &               &        2.80   &               &               &        2.15   \\
\multicolumn{1}{l}{K-P F-Stat FNC \& MSNBC (Joint)}&               &               &        8.57   &               &               &        9.55   \\

%% file: tables/first_stage_raw2.tex
\\ MSNBC Channel Position&      -0.026   &      -0.039***&      -0.044***&      -0.019   &      -0.037***&      -0.041***\\
            &     (0.018)   &     (0.011)   &     (0.010)   &     (0.022)   &     (0.011)   &     (0.011)   \\
[1ex] Fox News Channel Position&               &               &       0.035***&               &               &       0.007   \\
            &               &               &     (0.011)   &               &               &     (0.019)   \\

\\ \hline \multicolumn{1}{l}{Observations}&         718   &         718   &         718   &         718   &         718   &         718   \\
\multicolumn{1}{l}{K-P F-Stat FNC}&               &               &       10.63   &               &               &        0.13   \\
\multicolumn{1}{l}{K-P F-Stat MSNBC}&        2.11   &       11.55   &       18.17   &        0.73   &       10.84   &       13.01   \\
\multicolumn{1}{l}{K-P F-Stat FNC \& MSNBC (Joint)}&               &               &       14.66   &               &               &        7.02   \\

%% file: tables/identification_check.tex
\begin{table}[ht!]
\centering
\small
\begin{threeparttable}
\caption{Identification Checks: Fox News Position Correlation with Covariates \label{tab:identification-check}}
\begin{tabular}{l*{2}{r}} \hline
\noalign{\vspace{2ex}}
& \multicolumn{2}{c}{\textbf{Predictions for:}} \\ [1ex]
& \multicolumn{1}{c}{\textbf{}} & \multicolumn{1}{c}{\textbf{Republican}} \\
& \multicolumn{1}{c}{\textbf{Fox News Viewership}} & \multicolumn{1}{c}{\textbf{Newspaper Slant}} \\ [1ex]

& (1) & (2) \\
\hline \hline

\input{tables/identification_check_raw} \vspace{-2.8ex} \\
\midrule 
State Fixed Effects &           \checkmark  &          \checkmark   \\
Demographic Controls &           &              \\
Channel Controls &         &                \ \\
\bottomrule
\end{tabular}
\end{threeparttable}
\begin{minipage}{0.99\textwidth}
\vskip 0.2cm
{\footnotesize \emph{Notes:} Each column represents a separate regression. Observations are individual newspapers. The main explanatory variable is FNC's position (standardized). The outcomes are predictions: Fox News Channel viewership (column 1) and Republican newspaper slant (column 2). Both predicted outcomes are generated using demographic and channel system characteristics and state fixed effects. All specifications include state fixed effects. Appendix \ref{app:sec:sumstats} lists all control variables. We weight observations by circulation. We cluster standard errors at the state level (in parentheses; *** p<0.01, ** p<0.05, * p<0.1). \par}
\end{minipage}
\end{table}

%% file: tables/identification_check_raw.tex
\\ Fox News Channel Position&       0.006   &       0.001   \\
            &     (0.054)   &     (0.025)   \\

\\ \hline \multicolumn{1}{l}{Observations}&         718   &         718   \\

%% file: tables/ols.tex
\begin{table}[ht!]
\centering
\small
\begin{threeparttable}
\caption{OLS: Fox News Viewership Effects on Newspaper Slant \label{tab:ols}}
\begin{tabular}{l*{4}{r}} \hline
\noalign{\vspace{2ex}}
& \multicolumn{4}{c}{\textbf{Republican}} \\
& \multicolumn{4}{c}{\textbf{Newspaper Slant}} \\
& \multicolumn{4}{c}{\textbf{All Outlets}} \\ [1ex]
& (1) & (2) & (3) & (4) \\
\hline \hline

\input{tables/ols_raw1} \vspace{-2.8ex} \\
\midrule \\[1ex]

& \multicolumn{4}{c}{\textbf{Republican}} \\
& \multicolumn{4}{c}{\textbf{Newspaper Slant}} \\
& \multicolumn{4}{c}{\textbf{Outlets $>$ 1000 Articles}} \\ [1ex]
& (1) & (2) & (3) & (4) \\
\hline \hline

\input{tables/ols_raw2} \vspace{-2.8ex} \\ 
\midrule 
State Fixed Effects  &         \checkmark     &  \checkmark &  \checkmark &  \checkmark \\
Demographic Controls &           &         \checkmark   &         \checkmark &         \checkmark \\
Channel Controls &          &           &         \checkmark  &         \checkmark \\
Newspaper Language Controls &         &          &         &         \checkmark \\
\bottomrule
\end{tabular}
\end{threeparttable}
\begin{minipage}{0.99\textwidth}
\vskip 0.2cm
{\footnotesize \emph{Notes:} Each column represents a separate regression. Observations are individual newspapers. The upper panel includes all newspapers, while the lower panel restricts to newspapers with more than 1,000 articles. The main explanatory variable is Fox News channel viewership. The outcome is the newspaper's Republican slant (standardized), measured as the share of Republican-leaning articles among all articles. All specifications include state fixed effects. Demographic controls, channel controls, and generic language controls are introduced progressively in columns (2)-(4). Appendix \ref{app:sec:sumstats} lists all control variables. We weight observations by circulation. We cluster standard errors at the state level (in parentheses; *** p<0.01, ** p<0.05, * p<0.1). \par}
\end{minipage}
\end{table}

%% file: tables/ols_raw1.tex
\\ Fox News Channel Viewership&       0.128*  &       0.097   &       0.072   &       0.089   \\
            &     (0.074)   &     (0.079)   &     (0.083)   &     (0.068)   \\
[1ex] Households with FNC Access [Share]&               &               &       2.993   &       2.233   \\
            &               &               &     (2.567)   &     (1.943)   \\
[1ex] Households with MSNBC Access [Share]&               &               &      -2.634   &      -2.221*  \\
            &               &               &     (1.721)   &     (1.279)   \\

\\ \hline \multicolumn{1}{l}{Observations}&         718   &         718   &         718   &         718   \\

%% file: tables/ols_raw2.tex
\\ Fox News Channel Viewership&       0.180** &       0.157***&       0.146***&       0.125** \\
            &     (0.072)   &     (0.050)   &     (0.052)   &     (0.053)   \\
[1ex] Households with FNC Access [Share]&               &               &       0.964   &       0.756   \\
            &               &               &     (0.758)   &     (0.689)   \\
[1ex] Households with MSNBC Access [Share]&               &               &      -0.499   &      -0.344   \\
            &               &               &     (0.407)   &     (0.355)   \\

\\ \hline \multicolumn{1}{l}{Observations}&         541   &         541   &         541   &         541   \\

%% file: tables/reduced_form.tex
\begin{table}[ht!]
\centering
\small
\begin{threeparttable}
\caption{Reduced Form: Newspaper Slant and Fox News Position \label{tab:reduced_form}}
\begin{tabular}{l*{3}{r}} \hline
\noalign{\vspace{2ex}}
& \multicolumn{3}{c}{\textbf{Republican}} \\
& \multicolumn{3}{c}{\textbf{Newspaper Slant}} \\ [1ex]

& (1) & (2) & (3) \\
\hline \hline

\input{tables/reduced_form_raw} \vspace{-2.8ex} \\
\midrule 
State Fixed Effects   &         \checkmark     &  \checkmark &  \checkmark \\
Demographic Controls &         \checkmark   &         \checkmark  &         \checkmark \\
Channel Controls &            &         \checkmark   &         \checkmark \\
Newspaper Language Controls &            &        &         \checkmark \\
\bottomrule
\end{tabular}
\end{threeparttable}
\begin{minipage}{0.99\textwidth}
\vskip 0.2cm
{\footnotesize \emph{Notes:} Each column represents a separate regression. Observations are individual newspapers. All estimates are obtained using OLS. The main explanatory variable is the Fox News channel position (standardized). The outcome is the newspaper's Republican slant (standardized), measured as the share of Republican-leaning articles among all articles. All specifications include state fixed effects and demographic controls. Additionally, column (2) includes channel controls; column (3) further adds generic language controls. Appendix \ref{app:sec:sumstats} lists all control variables. We weight observations by circulation. We cluster standard errors at the state level (in parentheses; *** p<0.01, ** p<0.05, * p<0.1). \par}
\end{minipage}
\end{table}

%% file: tables/reduced_form_raw.tex
\\ Fox News Channel Position&      -0.111***&      -0.091***&      -0.094** \\
            &     (0.027)   &     (0.031)   &     (0.047)   \\
[1ex] MSNBC Channel Position&               &       0.057** &       0.018   \\
            &               &     (0.028)   &     (0.024)   \\

\\ \hline \multicolumn{1}{l}{Observations}&         718   &         718   &         718   \\

%% file: tables/main_msnbc.tex
\begin{table}[ht!]
\centering
\small
\begin{threeparttable}
\caption{2SLS: Newspaper Slant and Fox News Viewership \label{tab:main_msnbc}}
\begin{tabular}{l*{3}{r}} \hline
\noalign{\vspace{2ex}}
& \multicolumn{3}{c}{\textbf{Republican}} \\
& \multicolumn{3}{c}{\textbf{Newspaper Slant}} \\ [1ex]

& (1) & (2) & (3) \\
\hline \hline

\input{tables/main_msnbc_raw} \vspace{-2.8ex} \\
\midrule 
State Fixed Effects   &         \checkmark     &  \checkmark &  \checkmark \\
Demographic Controls &         \checkmark   &         \checkmark   &         \checkmark \\
Channel Controls &            &         \checkmark   &         \checkmark \\
Newspaper Language Controls &            &            &         \checkmark \\
\bottomrule
\end{tabular}
\end{threeparttable}
\begin{minipage}{0.99\textwidth}
\vskip 0.2cm
{\footnotesize \emph{Notes:} Each column represents a separate regression. Observations are individual newspapers. All estimates are obtained using two-stage least squares (2SLS). The main explanatory variable is MSNBC channel viewership instrumented with the channel's position (standardized). The outcome is the newspaper's Republican slant (standardized), measured as the share of Republican-leaning articles among all articles. All specifications include state fixed effects and demographic controls. Additionally, column (2) includes channel controls; column (3) further adds generic language controls. Appendix \ref{app:sec:sumstats} lists all control variables. We weight observations by circulation. We cluster standard errors at the state level (in parentheses; *** p<0.01, ** p<0.05, * p<0.1). \par}
\end{minipage}
\end{table}

%% file: tables/main_msnbc_raw.tex
\\ MSNBC Viewership&      -1.153   &      -1.399*  &      -0.420   \\
            &     (0.847)   &     (0.765)   &     (0.578)   \\

\\ \hline \multicolumn{1}{l}{Observations}&         718   &         718   &         718   \\
\multicolumn{1}{l}{K-P First-Stage F-Stat}&       10.84   &       13.01   &       14.13   \\

%% file: tables/main_rep_over_partisan.tex
\begin{table}[ht!]
\centering
\small
\begin{threeparttable}
\caption{\centering 2SLS: Newspaper Slant and Fox News Viewership\\
Robustness: Slant Only Relative to Partisan Articles\label{tab:rep_over_partisan}}
\begin{tabular}{l*{3}{r}} \hline
\noalign{\vspace{2ex}}
& \multicolumn{3}{c}{\textbf{Republican}} \\
& \multicolumn{3}{c}{\textbf{Newspaper Slant}} \\ 
& \multicolumn{3}{c}{\textbf{Partisan Articles Only}} \\ [1ex]
& (1) & (2) & (3) \\
\hline \hline

\input{tables/main_rep_over_partisan_raw} \vspace{-2.8ex} \\
\midrule 
State Fixed Effects   &         \checkmark     &  \checkmark &  \checkmark \\
Demographic Controls &         \checkmark   &         \checkmark   &         \checkmark \\
Channel Controls &            &         \checkmark   &         \checkmark \\
Newspaper Language Controls &          &            &         \checkmark \\
\bottomrule
\end{tabular}
\end{threeparttable}
\begin{minipage}{0.99\textwidth}
\vskip 0.2cm
{\footnotesize \emph{Notes:} Each column represents a separate regression. Observations are individual newspapers. All estimates are obtained using two-stage least squares (2SLS). The main explanatory variable is Fox News channel viewership instrumented with the channel's position (standardized). This table is a robustness check, and the outcome is the newspaper's Republican slant (standardized) but measured as the share of Republican-leaning articles among all partisan articles (instead of all articles as in Table \ref{tab:main_instr}). All specifications include state fixed effects and demographic controls. Additionally, column (2) includes channel controls; column (3) further adds generic language controls. Appendix \ref{app:sec:sumstats} lists all control variables. We weight observations by circulation. We cluster standard errors at the state level (in parentheses; *** p<0.01, ** p<0.05, * p<0.1). \par}
\end{minipage}
\end{table}

%% file: tables/main_rep_over_partisan_raw.tex
\\ Fox News Channel Viewership&       0.702** &       0.424** &       0.559***\\
            &     (0.317)   &     (0.204)   &     (0.207)   \\

\\ \hline \multicolumn{1}{l}{Observations}&         717   &         717   &         717   \\
\multicolumn{1}{l}{K-P First-Stage F-Stat}&        8.99   &       14.69   &       15.05   \\

%% file: tables/main_demdum.tex
\begin{table}[ht!]
\centering
\small
\begin{threeparttable}
\caption{\centering 2SLS: Newspaper Slant and Fox News Viewership\\
Robustness: Democrat Slant \label{tab:demdum}}
\begin{tabular}{l*{3}{r}} \hline
\noalign{\vspace{2ex}}
& \multicolumn{3}{c}{\textbf{Democrat}} \\
& \multicolumn{3}{c}{\textbf{Newspaper Slant}} \\ [1ex]

& (1) & (2) & (3) \\
\hline \hline

\input{tables/main_demdum_raw} \vspace{-2.8ex} \\
\midrule 
State Fixed Effects   &         \checkmark     &  \checkmark &  \checkmark \\
Demographic Controls &         \checkmark   &         \checkmark   &         \checkmark \\
Channel Controls &            &         \checkmark  &         \checkmark \\
Newspaper Language Controls &            &           &         \checkmark \\
\bottomrule
\end{tabular}
\end{threeparttable}
\begin{minipage}{0.99\textwidth}
\vskip 0.2cm
{\footnotesize \emph{Notes:} Each column represents a separate regression. Observations are individual newspapers. All estimates are obtained using two-stage least squares (2SLS). The main explanatory variable is Fox News channel viewership instrumented with the channel's position (standardized). This table is a robustness check, and the outcome is the newspaper's Democrat slant (standardized) (instead of its Republican slant as in Table \ref{tab:main_instr}). All specifications include state fixed effects and demographic controls. Additionally, column (2) includes channel controls; column (3) further adds generic language controls. Appendix \ref{app:sec:sumstats} lists all control variables. We weight observations by circulation. We cluster standard errors at the state level (in parentheses; *** p<0.01, ** p<0.05, * p<0.1). \par}
\end{minipage}
\end{table}

%% file: tables/main_demdum_raw.tex
\\ Fox News Channel Viewership&      -0.364   &      -0.019   &      -0.199   \\
            &     (0.347)   &     (0.185)   &     (0.185)   \\

\\ \hline \multicolumn{1}{l}{Observations}&         718   &         718   &         718   \\
\multicolumn{1}{l}{K-P First-Stage F-Stat}&        9.01   &       14.70   &       15.06   \\

%% file: tables/main_few1000.tex
\begin{table}[ht!]
\centering
\begin{threeparttable}
\caption{\centering 2SLS: Newspaper Slant and Fox News Viewership\\
Robustness: Outlets with $\geq$ 1000 Articles \label{tab:main_few1000}}
\small
\begin{tabular}{l*{3}{r}} \hline
\noalign{\vspace{2ex}}
& \multicolumn{3}{c}{\textbf{Republican}} \\ 
& \multicolumn{3}{c}{\textbf{Newspaper Slant}} \\ 
& \multicolumn{3}{c}{\textbf{Outlets $\geq$ 1000 Art.}} \\
[1ex]

& (1) & (2) & (3) \\
\hline \hline

\input{tables/main_few1000_raw} \vspace{-2.8ex} \\
\midrule 
State Fixed Effects  &         \checkmark     &  \checkmark &  \checkmark \\
Demographic Controls &         \checkmark   &         \checkmark  &         \checkmark \\
Channel Controls &           &         \checkmark   &         \checkmark \\
Newspaper Language Controls &           &           &         \checkmark \\
\bottomrule
\end{tabular}
\end{threeparttable}
\begin{minipage}{0.99\textwidth}
\vskip 0.2cm
{\footnotesize \emph{Notes:} Each column represents a separate regression. Observations are individual newspapers. This is a robustness check, and we drop newspapers with less than 1000 articles in 2005-08. All estimates are obtained using two-stage least squares (2SLS). The main explanatory variable is Fox News channel viewership instrumented with the channel's position (standardized). The outcome is the newspaper's Republican slant (standardized), measured as the share of Republican-leaning articles among all articles. All specifications include state fixed effects and demographic controls. Additionally, column (2) includes channel controls; column (3) further adds generic language controls. Appendix \ref{app:sec:sumstats} lists all control variables. We weight observations by circulation. We cluster standard errors at the state level (in parentheses; *** p<0.01, ** p<0.05, * p<0.1). \par}
\end{minipage}
\end{table}

%% file: tables/main_few1000_raw.tex
\\ Fox News Channel Viewership&       0.456*  &       0.550** &       0.481*  \\
            &     (0.249)   &     (0.236)   &     (0.262)   \\

\\ \hline \multicolumn{1}{l}{Observations}&         541   &         541   &         541   \\
\multicolumn{1}{l}{K-P First-Stage F-Stat}&        8.78   &       12.13   &       11.53   \\

%% file: tables/main_keepobi.tex
\begin{table}[ht!]
\centering
\begin{threeparttable}
\caption{\centering 2SLS: Newspaper Slant and Fox News Viewership\\
Robustness: Keeping Routine Announcements (e.g., Obituaries) \label{tab:main_keepobi}}
\small
\begin{tabular}{l*{3}{r}} \hline
\noalign{\vspace{2ex}}
& \multicolumn{3}{c}{\textbf{Republican}} \\
& \multicolumn{3}{c}{\textbf{Newspaper Slant}} \\
& \multicolumn{3}{c}{\textbf{All Articles}} \\ [1ex]

& (1) & (2) & (3) \\
\hline \hline

\input{tables/main_keepobi_raw.tex} \vspace{-2.8ex} \\
\midrule 
State Fixed Effects   &         \checkmark     &  \checkmark &  \checkmark \\
Demographic Controls &         \checkmark   &         \checkmark   &         \checkmark \\
Channel Controls &          &         \checkmark   &         \checkmark \\
Newspaper Language Controls &        &           &         \checkmark \\
\bottomrule
\end{tabular}
\end{threeparttable}
\begin{minipage}{0.99\textwidth}
\vskip 0.2cm
{\footnotesize \emph{Notes:} Each column represents a separate regression. Observations are individual newspapers. All estimates are obtained using two-stage least squares (2SLS). The main explanatory variable is Fox News channel viewership instrumented with the channel's position (standardized). The outcome is the newspaper's Republican slant (standardized), measured as the share of Republican-leaning articles among all articles. This is a robustness check, and we do not drop routine announcements (e.g., death notices) before aggregating slant at the newspaper level (as we do in Table \ref{tab:main_instr}). All specifications include state fixed effects and demographic controls. Additionally, column (2) includes channel controls; column (3) further adds generic language controls. Appendix \ref{app:sec:sumstats} lists all control variables. We weight observations by circulation. We cluster standard errors at the state level (in parentheses; *** p<0.01, ** p<0.05, * p<0.1). \par}
\end{minipage}
\end{table}

%% file: tables/main_keepobi_raw.tex
\\ Fox News Channel Viewership&       0.596** &       0.450***&       0.470** \\
            &     (0.261)   &     (0.164)   &     (0.196)   \\

\\ \hline \multicolumn{1}{l}{Observations}&         718   &         718   &         718   \\
\multicolumn{1}{l}{K-P First-Stage F-Stat}&        9.01   &       14.70   &       15.06   \\

%% file: tables/main_weights_hist.tex
\begin{table}[ht!]
\centering
\small
\begin{threeparttable}
\caption{\centering 2SLS: Newspaper Slant and Fox News Viewership\\
Robustness: Historical Circulation \label{tab:main_weights_hist}}
\begin{tabular}{l*{3}{r}} \hline
\noalign{\vspace{2ex}}
& \multicolumn{3}{c}{\textbf{Republican}} \\
& \multicolumn{3}{c}{\textbf{Newspaper Slant}} \\ [1ex]

& (1) & (2) & (3) \\
\hline \hline

\input{tables/main_weights_hist_raw} \vspace{-2.8ex} \\
\midrule 
State Fixed Effects   &         \checkmark     &  \checkmark &  \checkmark \\
Demographic Controls &         \checkmark  &         \checkmark   &         \checkmark\\
Channel Controls &            &         \checkmark   &         \checkmark \\
Newspaper Language Controls &            &         &         \checkmark \\
\bottomrule
\end{tabular}
\end{threeparttable}
\begin{minipage}{0.99\textwidth}
\vskip 0.2cm
{\footnotesize \emph{Notes:} Each column represents a separate regression. Observations are individual newspapers. All estimates are obtained using two-stage least squares (2SLS). The main explanatory variable is Fox News channel viewership instrumented with the channel's position (standardized). The outcome is the newspaper's Republican slant (standardized), measured as the share of Republican-leaning articles among all articles. All specifications include state fixed effects and demographic controls. Additionally, column (2) includes channel controls; column (3) further adds generic language controls. Appendix \ref{app:sec:sumstats} lists all control variables. This is a robustness check of Table \ref{tab:main_instr}, and we weight observations by 1996 circulation here (instead of contemporary circulation for 2005-08). We cluster standard errors at the state level (in parentheses; *** p<0.01, ** p<0.05, * p<0.1). \par}
\end{minipage}
\end{table}

%% file: tables/main_weights_hist_raw.tex
\\ Fox News Channel Viewership&       0.831*  &       0.461*  &       0.449*  \\
            &     (0.445)   &     (0.243)   &     (0.229)   \\

\\ \hline \multicolumn{1}{l}{Observations}&         608   &         608   &         608   \\
\multicolumn{1}{l}{K-P First-Stage F-Stat}&        5.67   &       10.11   &       10.35   \\

%% file: tables/main_weights_rel.tex
\begin{table}[ht!]
\centering
\begin{threeparttable}
\caption{\centering 2SLS: Newspaper Slant and Fox News Viewership\\
Robustness: Relative Circulation\label{tab:main_weights_rel}}
\small
\begin{tabular}{l*{3}{r}} \hline
\noalign{\vspace{2ex}}
& \multicolumn{3}{c}{\textbf{Republican}} \\
& \multicolumn{3}{c}{\textbf{Newspaper Slant}} \\ [1ex]

& (1) & (2) & (3) \\
\hline \hline

\input{tables/main_weights_rel_raw} \vspace{-2.8ex} \\
\midrule 
State Fixed Effects   &         \checkmark    &  \checkmark &  \checkmark \\
Demographic Controls &         \checkmark   &         \checkmark  &         \checkmark \\
Channel Controls &           &         \checkmark   &         \checkmark \\
Newspaper Language Controls &          &           &         \checkmark \\
\bottomrule
\end{tabular}
\end{threeparttable}
\begin{minipage}{0.99\textwidth}
\vskip 0.2cm
{\footnotesize \emph{Notes:} Each column represents a separate regression. Observations are individual newspapers. All estimates are obtained using two-stage least squares (2SLS). The main explanatory variable is Fox News channel viewership instrumented with the channel's position (standardized). The outcome is the newspaper's Republican slant (standardized), measured as the share of Republican-leaning articles among all articles. All specifications include state fixed effects and demographic controls. Additionally, column (2) includes channel controls; column (3) further adds generic language controls. Appendix \ref{app:sec:sumstats} lists all control variables. This is a robustness check of Table \ref{tab:main_instr}, and we weight observations by their relative circulation in the county here (instead of absolute newspaper-level circulation). We cluster standard errors at the state level (in parentheses; *** p<0.01, ** p<0.05, * p<0.1). \par}
\end{minipage}
\end{table}

%% file: tables/main_weights_rel_raw.tex
\\ Fox News Channel Viewership&       0.707** &       0.591** &       0.625***\\
            &     (0.322)   &     (0.222)   &     (0.229)   \\

\\ \hline \multicolumn{1}{l}{Observations}&         718   &         718   &         718   \\
\multicolumn{1}{l}{K-P First-Stage F-Stat}&        8.90   &       17.50   &       17.61   \\

%% file: tables/main_fncmsnbc.tex
\begin{table}[ht!]
\centering
\begin{threeparttable}
\caption{\centering 2SLS: Newspaper Slant and Fox News Viewership\\
Robustness: Fox News Relative to MSNBC Viewership \label{tab:main_fncmsnbc}}
\small
\begin{tabular}{l*{3}{r}} \hline
\noalign{\vspace{2ex}}
& \multicolumn{3}{c}{\textbf{Republican}} \\
& \multicolumn{3}{c}{\textbf{Newspaper Slant}} \\ [1ex]

& (1) & (2) & (3) \\
\hline \hline

\input{tables/main_fncmsnbc_raw} \vspace{-2.8ex} \\
\midrule 
State Fixed Effects   &         \checkmark     &  \checkmark &  \checkmark \\
Demographic Controls &         \checkmark  &         \checkmark   &         \checkmark\\
Channel Controls &           &         \checkmark   &         \checkmark \\
Newspaper Language Controls &            &          &         \checkmark \\
\bottomrule
\end{tabular}
\end{threeparttable}
\begin{minipage}{0.99\textwidth}
\vskip 0.2cm
{\footnotesize \emph{Notes:} Each column represents a separate regression. Observations are individual newspapers. All estimates are obtained using two-stage least squares (2SLS). The main explanatory variable is the Fox News channel viewership relative to the MSNBC viewership instrumented with the relative distance between the Fox News and the MSNBC channel positions (standardized). The outcome is the newspaper's Republican slant (standardized), measured as the share of Republican-leaning articles among all articles. All specifications include state fixed effects and demographic controls. Additionally, column (2) includes channel controls; column (3) further adds generic language controls. Appendix \ref{app:sec:sumstats} lists all control variables. We weight observations by circulation. We cluster standard errors at the state level (in parentheses; *** p<0.01, ** p<0.05, * p<0.1). \par}
\end{minipage}
\end{table}

%% file: tables/main_fncmsnbc_raw.tex
\\ FNC Viewership relative to MSNBC&       1.085***&       0.875***&       0.916** \\
            &     (0.384)   &     (0.305)   &     (0.411)   \\

\\ \hline \multicolumn{1}{l}{Observations}&         718   &         718   &         718   \\
\multicolumn{1}{l}{K-P First-Stage F-Stat}&       16.96   &       10.72   &       10.66   \\

%% file: tables/main_dropintab.tex
\begin{table}[ht!]
\centering
\begin{threeparttable}
\caption{\centering 2SLS: Newspaper Slant and Fox News Viewership\\
Robustness: Dropping Counties with Low Nielsen Respondent Number \label{tab:main_dropintab}}
\small
\begin{tabular}{l*{3}{r}} \hline
\noalign{\vspace{2ex}}
& \multicolumn{3}{c}{\textbf{Republican}} \\
& \multicolumn{3}{c}{\textbf{Newspaper Slant}} \\ [1ex]

& (1) & (2) & (3) \\
\hline \hline

\input{tables/main_dropintab_raw.tex} \vspace{-2.8ex} \\
\midrule 
State Fixed Effects   &         \checkmark     &  \checkmark &  \checkmark \\
Demographic Controls &         \checkmark   &         \checkmark   &         \checkmark\\
Channel Controls &            &         \checkmark   &         \checkmark \\
Newspaper Language Controls &           &         &         \checkmark \\
\bottomrule
\end{tabular}
\end{threeparttable}
\begin{minipage}{0.99\textwidth}
\vskip 0.2cm
{\footnotesize \emph{Notes:} Each column represents a separate regression. Observations are individual newspapers. This is a robustness check of Table \ref{tab:main_instr} where we drop counties with few Nielsen survey respondents (below the 25th percentile). All estimates are obtained using two-stage least squares (2SLS). The main explanatory variable is Fox News channel viewership instrumented with the channel's position (standardized). The outcome is the newspaper's Republican slant (standardized), measured as the share of Republican-leaning articles among all articles. All specifications include state fixed effects and demographic controls. Additionally, column (2) includes channel controls; column (3) further adds generic language controls. Appendix \ref{app:sec:sumstats} lists all control variables. We weight observations by circulation. We cluster standard errors at the state level (in parentheses; *** p<0.01, ** p<0.05, * p<0.1). \par}
\end{minipage}
\end{table}

%% file: tables/main_dropintab_raw.tex
\\ Fox News Channel Viewership&       0.776** &       0.656***&       0.657***\\
            &     (0.315)   &     (0.221)   &     (0.216)   \\

\\ \hline \multicolumn{1}{l}{Observations}&         558   &         558   &         558   \\
\multicolumn{1}{l}{K-P First-Stage F-Stat}&        7.38   &       13.96   &       14.50   \\

%% file: tables/main_censusdiv.tex
\begin{table}[ht!]
\centering
\begin{threeparttable}
\caption{\centering 2SLS: Newspaper Slant and Fox News Viewership\\
Robustness: Census Division Fixed Effects \label{tab:main_censusdiv}}
\small
\begin{tabular}{l*{3}{r}} \hline
\noalign{\vspace{2ex}}
& \multicolumn{3}{c}{\textbf{Republican}} \\
& \multicolumn{3}{c}{\textbf{Newspaper Slant}} \\ [1ex]

& (1) & (2) & (3) \\
\hline \hline

\input{tables/main_censusdiv_raw} \vspace{-2.8ex} \\
\midrule 
Census Division Fixed Effects   &         \checkmark     &  \checkmark &  \checkmark \\
Demographic Controls &         \checkmark   &         \checkmark   &         \checkmark \\
Channel Controls &           &         \checkmark   &         \checkmark \\
Newspaper Language Controls &           &           &         \checkmark \\
\bottomrule
\end{tabular}
\end{threeparttable}
\begin{minipage}{0.99\textwidth}
\vskip 0.2cm
{\footnotesize \emph{Notes:} Each column represents a separate regression. Observations are individual newspapers. All estimates are obtained using two-stage least squares (2SLS). The main explanatory variable is Fox News channel viewership instrumented with the channel's position (standardized). The outcome is the newspaper's Republican slant (standardized), measured as the share of Republican-leaning articles among all articles. All specifications include census division fixed effects (as a robustness check of Table \ref{tab:main_instr} where we use state fixed effects) and demographic controls. Additionally, column (2) includes channel controls; column (3) further adds generic language controls. Appendix \ref{app:sec:sumstats} lists all control variables. We weight observations by circulation. We cluster standard errors at the state level (in parentheses; *** p<0.01, ** p<0.05, * p<0.1). \par}
\end{minipage}
\end{table}

%% file: tables/main_censusdiv_raw.tex
\\ Fox News Channel Viewership&       0.782***&       0.659***&       0.467** \\
            &     (0.267)   &     (0.196)   &     (0.178)   \\

\\ \hline \multicolumn{1}{l}{Observations}&         718   &         718   &         718   \\
\multicolumn{1}{l}{K-P First-Stage F-Stat}&        9.51   &       15.61   &       17.07   \\

%% file: tables/main_instrnoclust.tex
\begin{table}[ht!]
\centering
\small
\begin{threeparttable}
\caption{\centering2SLS: Newspaper Slant and Fox News Viewership \\
Robustness: Robust Standard Errors \label{tab:main_instrnoclust}}
\begin{tabular}{l*{3}{r}} \hline
\noalign{\vspace{2ex}}
& \multicolumn{3}{c}{\textbf{Newspaper Slant}} \\ [1ex]

& (1) & (2) & (3) \\
\hline \hline

\input{tables/main_instrnoclust_raw} \vspace{-2.8ex} \\
\midrule 
State Fixed Effects   &         \checkmark     &  \checkmark &  \checkmark \\
Demographic Controls &         \checkmark   &         \checkmark  &         \checkmark \\
Channel Controls &            &         \checkmark   &         \checkmark \\
Newspaper Language Controls &            &           &         \checkmark \\
\bottomrule
\end{tabular}
\end{threeparttable}
\begin{minipage}{0.99\textwidth}
\vskip 0.2cm
{\footnotesize \emph{Notes:} Each column represents a separate regression. Observations are individual newspapers. All estimates are obtained using two-stage least squares (2SLS). The main explanatory variable is Fox News channel viewership instrumented with the channel's position (standardized). The outcome is the newspaper's Republican slant (standardized), measured as the share of Republican-leaning articles among all articles. All specifications include state fixed effects and demographic controls. Additionally, column (2) includes channel controls; column (3) further adds generic language controls. Appendix \ref{app:sec:sumstats} lists all control variables. We weight observations by circulation. We report robust standard errors (in parentheses; *** p<0.01, ** p<0.05, * p<0.1). \par}
\end{minipage}
\end{table}

%% file: tables/main_instrnoclust_raw.tex
\\ Fox News Channel Viewership&       0.718*  &       0.493*  &       0.515** \\
            &     (0.368)   &     (0.267)   &     (0.257)   \\

\\ \hline \multicolumn{1}{l}{Observations}&         718   &         718   &         718   \\

%% file: tables/top_covariates.tex
\begin{tabular}{ll}
\hline
\textbf{Strong Response of Slant} & \textbf{Strong Response of Voting} \\
\hline
Share age 30-59 & Share males \\
Share Hispanics & Share age 30-59 \\
Share service occupations & Share arts occupations \\
Share office occupations & Median tax rate \\
\hline
\end{tabular}

%% file: tables/mech_local.tex
\begin{table}[ht!]
\centering
\begin{threeparttable}
\caption{\centering 2SLS: Newspaper Slant and Fox News Viewership\\
Mechanisms: Slant in Local News Articles\label{tab:mech_local}}
\small
\begin{tabular}{l*{4}{r}} \hline
\noalign{\vspace{2ex}}
& & \multicolumn{3}{c}{\textbf{Republican}} \\
& \textbf{Local} & \multicolumn{3}{c}{\textbf{Newspaper Slant}} \\
& \textbf{Share} & \multicolumn{3}{c}{\textbf{Local Articles}} \\ [1ex]
& (1) & (2) & (3) & (4) \\
\hline \hline

\input{tables/mech_local_raw} \vspace{-2.8ex} \\
\midrule 
State Fixed Effects   &         \checkmark     &  \checkmark &  \checkmark &  \checkmark \\
Demographic Controls &         \checkmark   &         \checkmark   &         \checkmark &  \checkmark \\
Channel Controls &         \checkmark   &           &         \checkmark & \checkmark \\
Newspaper Language Controls &         \checkmark   &        &         &    \checkmark   \\
\bottomrule
\end{tabular}
\end{threeparttable}
\begin{minipage}{0.99\textwidth}
\vskip 0.2cm
{\footnotesize \emph{Notes:} Each column represents a separate regression. Observations are individual newspapers. All estimates are obtained using two-stage least squares (2SLS). The main explanatory variable is Fox News channel viewership instrumented with the channel's position (standardized). In the first column, the outcome is the share of articles covering local news (rather than national or international news). In columns (2) to (4), the outcome is the newspaper's Republican slant (standardized), measured as the share of Republican-leaning local articles among all local articles. This is a robustness check of Table \ref{tab:main_instr}. Here, we only consider articles covering local news. All specifications include state fixed effects and demographic controls. Channel controls are included in columns (1), (3) and (4); generic language controls in columns (1) and (4). Appendix \ref{app:sec:sumstats} lists all control variables. We weight observations by circulation. We cluster standard errors at the state level (in parentheses; *** p<0.01, ** p<0.05, * p<0.1). \par}
\end{minipage}
\end{table}

%% file: tables/mech_local_raw.tex
\\ Fox News Channel Viewership&      -0.045   &       0.628** &       0.381** &       0.423** \\
            &     (0.038)   &     (0.270)   &     (0.150)   &     (0.185)   \\

\\ \hline \multicolumn{1}{l}{Observations}&         718   &         718   &         718   &         718   \\
\multicolumn{1}{l}{K-P First-Stage F-Stat}&       15.06   &       10.59   &       14.70   &       15.06   \\

%% file: tables/reduced_form_circ.tex
\begin{table}[ht!]
\centering
\small
\begin{threeparttable}
\caption{\centering Reduced Form: Circulation and Fox News Position\\
Mechanisms: Effects on Circulation \label{tab:reduced_form_circ}}
\begin{tabular}{l*{3}{r}} \hline
\noalign{\vspace{2ex}}
& \multicolumn{3}{c}{\textbf{Newspaper}} \\
& \multicolumn{3}{c}{\textbf{Circulation}} \\ [1ex]

& (1) & (2) & (3) \\
\hline \hline

\input{tables/reduced_form_circ_raw1} \vspace{-2.8ex} \\
\midrule 
State Fixed Effects   &         \checkmark     &  \checkmark &  \checkmark \\
Demographic Controls &         \checkmark   &         \checkmark  &         \checkmark \\
Channel Controls &           &         \checkmark   &         \checkmark \\
Newspaper Language Controls &         &            &         \checkmark \\
Historical Circulation Controls &          &           &         \\
\bottomrule
\end{tabular}
\end{threeparttable}
\begin{minipage}{0.99\textwidth}
\vskip 0.2cm
{\footnotesize \emph{Notes:} Each column represents a separate regression. Observations are individual newspapers. All estimates are obtained using Poisson Pseudo Maximum Likelihood (PPML) estimation. The main explanatory variable is the Fox News channel position (standardized). The outcome is the newspaper's circulation. All specifications include state fixed effects and demographic controls. Additionally, column (2) includes channel controls; column (3) further adds generic language controls. Appendix \ref{app:sec:sumstats} lists all control variables. We cluster standard errors at the state level (in parentheses; *** p<0.01, ** p<0.05, * p<0.1). \par}
\end{minipage}
\end{table}

%% file: tables/reduced_form_circ_raw1.tex
\\ Fox News Channel Position&      -0.011   &       0.013   &       0.009   \\
            &     (0.063)   &     (0.061)   &     (0.062)   \\
[1ex] MSNBC Channel Position&               &       0.002   &       0.008   \\
            &               &     (0.034)   &     (0.030)   \\

\\ \hline \multicolumn{1}{l}{Observations}&         718   &         718   &         718   \\

%% file: tables/reduced_form_circ_controlhist.tex
\begin{table}[ht!]
\centering
\small
\begin{threeparttable}
\caption{\centering Reduced Form: Circulation and Fox News Position\\
Mechanisms: Effects on Circulation \label{tab:reduced_form_circ_hist}}
\begin{tabular}{l*{3}{r}} \hline
\noalign{\vspace{2ex}}
& \multicolumn{3}{c}{\textbf{Newspaper}} \\
& \multicolumn{3}{c}{\textbf{Circulation}} \\ [1ex]

& (1) & (2) & (3) \\
\hline \hline

\input{tables/reduced_form_circ_raw2} \vspace{-2.8ex} \\
\midrule 
State Fixed Effects   &         \checkmark    &  \checkmark &  \checkmark \\
Demographic Controls &         \checkmark   &         \checkmark   &         \checkmark \\
Channel Controls &         &         \checkmark  &         \checkmark \\
Newspaper Language Controls &            &          &         \checkmark \\
Historical Circulation Controls &         \checkmark   &         \checkmark   &         \checkmark \\
\bottomrule
\end{tabular}
\end{threeparttable}
\begin{minipage}{0.99\textwidth}
\vskip 0.2cm
{\footnotesize \emph{Notes:} Each column represents a separate regression. Observations are individual newspapers. All estimates are obtained using Poisson Pseudo Maximum Likelihood (PPML) estimation. The main explanatory variable is the Fox News channel position (standardized). The outcome is the newspaper's circulation. All specifications include state fixed effects, demographic controls, and circulation before the Fox News channel's entry (in 1996). Additionally, column (2) includes channel controls; column (3) further adds generic language controls. Appendix \ref{app:sec:sumstats} lists all control variables. We cluster standard errors at the state level (in parentheses; *** p<0.01, ** p<0.05, * p<0.1). \par}
\end{minipage}
\end{table}

%% file: tables/reduced_form_circ_raw2.tex
\\ Fox News Channel Position&       0.011   &       0.013   &       0.012   \\
            &     (0.030)   &     (0.033)   &     (0.033)   \\
[1ex] MSNBC Channel Position&               &      -0.014   &      -0.014   \\
            &               &     (0.035)   &     (0.036)   \\

\\ \hline \multicolumn{1}{l}{Observations}&         608   &         608   &         608   \\